\documentclass[10pt,twocolumn] {IEEEtran}
\newtheorem{Remark}{\it Remark}[section]

\newtheorem{Proposition}{\it Proposition}[section]

\newtheorem{Definition}{\it Definition}[section]

\makeatletter
\newcommand{\Rmnum}[1]{\expandafter\@slowromancap\romannumeral #1@}
\makeatother
\usepackage{graphicx}
\usepackage{multicol}
\usepackage{setspace}
\usepackage{mathtools}
\usepackage{bm}
\usepackage{amsmath}
\usepackage{setspace}
\usepackage{enumerate}
\usepackage{caption}
\usepackage{epstopdf}
\usepackage{fancyhdr} 
\usepackage{indentfirst}
\usepackage{enumerate}
\usepackage{amssymb}
\usepackage{stfloats}
\usepackage{bm}
\usepackage{multirow}
\usepackage{threeparttable}
\usepackage{CJK}
\usepackage[justification=centering]{caption}
\usepackage{makecell}
\usepackage{caption}
\usepackage{cases}
\usepackage{amsmath}
\usepackage{graphics}
\usepackage{algorithm}
\usepackage{algpseudocode}

\usepackage{epsfig}
\usepackage{cite}
\usepackage{subfigure}
\usepackage{booktabs} 
\usepackage{color}
\usepackage{textcomp}

\begin{document}

\title{Receiver Selection and Transmit Beamforming for Multi-Static Integrated Sensing and Communications}
\author{
\IEEEauthorblockN{Dan Wang, Yuanming Tian, Chuan Huang, \textit{Member}, \textit{IEEE}, Hao Chen, Xiaodong Xu, \textit{Senior Member}, \textit{IEEE}, and Ping Zhang, \textit{Fellow}, \textit{IEEE}}
\thanks{
A conference version of this work \cite{conf_ISAC} was accepted by IEEE PIMRC Workshops 2024. \textit{Corresponding authors: Chuan Huang and Hao Chen.}

D. Wang and H. Chen are with the Department of of Broadband Communications, Peng Cheng Laboratory, Shenzhen, China, 518055. Emails: wangd01@pcl.ac.cn, chenh03@pcl.ac.cn.

Y. Tian is with the Shenzhen Future Network of Intelligence Institute, the School of Science and Engineering, and the Guangdong Provincial Key Laboratory of Future Networks of Intelligence, The Chinese University of Hong Kong, Shenzhen, China, 518172. Email: yuanmingtian@link.cuhk.edu.cn.

C. Huang is with the School of Science and Engineering, the Shenzhen Future Network of Intelligence Institute, and the Guangdong Provincial Key Laboratory of Future Networks of Intelligence, The Chinese University of Hong Kong, Shenzhen, China, 518172. Email: huangchuan@cuhk.edu.cn.

X. Xu and P. Zhang are with the State Key Laboratory of Networking and Switching Technology, Beijing University of Posts and Telecommunications, Beijing, China, 100876, and also with the Department of Broadband Communication, Peng Cheng Laboratory, Shenzhen, China, 518055. Emails: xuxiaodong@bupt.edu.cn, zhangping@bupt.edu.cn.

%

}
}
\maketitle
\begin{abstract}
Next-generation wireless networks are expected to develop a novel paradigm of integrated sensing and communications (ISAC) to enable both the high-accuracy sensing and high-speed communications. However, conventional mono-static ISAC systems, which simultaneously transmit and receive at the same equipment,
may suffer from severe self-interference, and thus significantly degrade the system performance.           
To address this issue, this paper studies a multi-static ISAC system for cooperative target localization and 
 communications, where the transmitter transmits ISAC signal to multiple receivers (REs) deployed at different positions. 
 We derive the closed-form Cram\'{e}r-Rao bound (CRB) on the joint estimations of both the transmission delay and Doppler shift for cooperative target localization, and the CRB minimization problem is formulated by considering the cooperative cost and communication rate requirements for the REs.
 To solve this problem, we first decouple it into two subproblems for RE selection and transmit beamforming, respectively. Then, a minimax linkage-based method is proposed to solve the RE selection subproblem, and a successive convex approximation algorithm is adopted to deal with the transmit beamforming subproblem with non-convex constraints.  
Finally, numerical results validate our analysis and reveal that our proposed multi-static ISAC scheme achieves better ISAC performance than the conventional mono-static ones with ideal SI cancellation when the number of cooperative REs is large.
%
%
\end{abstract}
\begin{IEEEkeywords}
Integrated sensing and communications (ISAC), multi-static, Cram\'{e}r-Rao bound (CRB), receiver (RE) selection, transmit beamforming, localization.
\end{IEEEkeywords}

\section{Introduction}
Next-generation wireless networks are expected to extend their capabilities from communication-only to environmental sensing to support various intelligent applications such as smart factories, vehicle-to-everything (V2X) communications, and remote healthcare \cite{10247107,10014666}. To improve the spectrum, energy, and hardware efficiency,  
integrated sensing and communications (ISAC) has been regarded as one of the key technologies to achieve both the sensing and communication functions by sharing the spectrum resource and hardware platforms \cite{10077112}. However, the inherent differences in tasks, functions and ways of working between wireless sensing and communications pose new design challenges for the ISAC systems.

%

The idea of ISAC originated from the coexistence of the sensing (radar) and communications (CSC) \cite{9582836} and the dual-functional sensing (radar)-communications (DFSC) \cite{9724174}. Specifically, CSC aimed to achieve the coexistence of two independent sensing and communication systems by using mutual interference management \cite{9442799,7089157,8355705}.
Following the concept of cognitive radio, the authors in \cite{9442799} proposed a spectrum sharing scheme between the sensing and communication in CSC, where the communication system operates when the spectrum is not occupied by the sensing system. 
The authors in \cite{7089157} proposed a null-space projection method to cancel the sensing interference for protecting the communication task, while it may result in poor sensing performance. To investigate the performance trade-off between the sensing and communication in CSC, the authors in \cite{8355705} utilized power control to enable the  multi-input-multi-output sensing and downlink multi-user multi-input-single-output communications by interference management.

On the other hand, DFSC focused on developing dual-functional systems to perform both the sensing and communication functions at the same time, as the joint design is potential to bring extra cooperation gain for the two systems \cite{9091840,9468975,10017835,8288677}.
By using the shared hardware, the targets to be sensed are viewed as virtual users, and thus various multiple access based methods, e.g., time-division multiple access \cite{9091840}, orthogonal frequency-division multiple access \cite{9468975}, and code-division multiple access \cite{10017835}, were adopted to realize DFSC by sharing wireless resource between the virtual users for sensing and the real ones for communications. 
 Moreover, spatial beamforming by using multiple antennas is another way to realize DFSC by exploiting the spatial degree of freedom (DoF). The authors in \cite{8288677} proposed a transmit beamforming scheme to achieve accuracy sensing with the communication constraints for both the separated and shared antenna deployments, and revealed that the shared deployment performs better than the separated one.  
To achieve better dual-functional performance, both the dedicated sensing and communication signals were utilized  in ISAC systems \cite{10086626}, and the ISAC system were shown to be beneficial from the dedicated sensing signal if the communication user are capable to cancel the interference from the sensing signals.


All the above works only focused on the mono-static paradigm, i.e., to transmit the ISAC signal and to receive the sensing echo at the same equipment, which results in severe self-interference (SI) \cite{Coordinated_Cellular}.
Moreover, sensing echo may return to the tranceiver equipment before the completion of the ISAC signal transmissions due to the experienced round-trip path loss, making it much weaker than the SI \cite{10283659}.
%
%
%
Hence, conventional SI cancellation techniques \cite{8647900, 9685832} designed for full-duplex communication systems cannot 
effectively suppress the SI for the mono-static ISAC systems.
To address this issue, recent researches have incorporated  bi-/multi-static paradigm into ISAC  \cite{10281382, Andrew_zhang, 10304081}, where the ISAC signal transmission and sensing echo reception are performed at differently located equipments to effectively avoid the extremely strong SI in the mono-static ISAC systems. 
Specifically, the authors in \cite{10281382} minimized the symbol error rate (SER) by establishing two Cram\'{e}r-Rao bound (CRB)  constraints on the direction of departure (DoD) and direction of arrival (DoA) in a bi-static ISAC system, which exhibited strong efficacy regarding the SER and resilience against the nonlinear phase noise compared to the conventional mono-static one. 
To enhance the sensing accuracy, the authors in \cite{Andrew_zhang} investigated the performance bound for Doppler estimation in the bi-static ISAC systems by optimizing the waveform for noise- and interference-limited sensing scenarios. 
Furthermore, the sensing coverage capability was investigated in \cite{10304081} for the cellular multi-static ISAC systems by maximizing the worst-case sensing signal-to-noise ratio in the prescribed sensing  region, with the minimum signal-to-interference-plus-noise ratio  requirement for each communication user.

However, most of the existing works concentrated on analyzing the performance bounds for the mono-/bi-static ISAC systems, which is not thoroughly studied for the multi-static scenarios.
As such, this paper studies a multi-static ISAC system, which includes one transmitter (TR), one target, and multiple receivers (REs) to simultaneously facilitate target localization and communications. Unlike the conventional mono-static one, the considered multi-static ISAC allows multiple REs to exchange their obtained transmission delay and Doppler shift information about the received signal and then cooperatively localize the target. By using multiple REs, space diversity can be exploited to improve the localization performance.
%
%
%
The main contributions of this work are summarized as follows:
\begin{itemize}
	\item 
First, we derive the transmitted and received ISAC signal models at the TR and multiple REs, respectively. Based on the derived signal model, we  analyze the joint estimations of transmission delay and Doppler shift at multiple REs for cooperative target localization, and derive the corresponding CRB in closed form. 
	\item  Then, a CRB minimization problem is formulated by considering the cooperative cost and communication rate requirements for these REs. To solve this problem, we decouple it into two sub-problems for RE selection and transmit beamforming, and then a minimax linkage-based RE selection method and a successive 
 convex approximation algorithm are proposed to solve these sub-problems, respectively.
	\item Finally, we model the Doppler shift of each TR-target-RE link as a two-step relativistic effect, and then derive the closed-form of DoA and distance between target and RE to
	localize the target by utilizing the transmission delay and Doppler shift information at multiple cooperative REs.
	\end{itemize}
	
	The remainder of this paper is organized as follows. Section \uppercase\expandafter{\romannumeral2} introduces the system model. Section \uppercase\expandafter{\romannumeral3} analyzes the performance of localization and communications, respectively, and then formulates the CRB minimization problem.
	 Section \uppercase\expandafter{\romannumeral4} 
	 proposes algorithms to solve the CRB minimization problem. Section \uppercase\expandafter{\romannumeral5} introduces a practical method to realize target localization. Section \uppercase\expandafter{\romannumeral6} 
	presents the numerical and simulation results. Section \uppercase\expandafter{\romannumeral7} concludes this paper.
	
	Notations:  Bold-face upper-case and lower-case letters, e.g.,  $\bf X$ and $\bf x$, denote matrices and vectors, respectively.
${\bf x}^T $ and ${\bf x}^H$ denote the transpose and
conjugate transpose of vector $\bf x$, respectively. ${\bf X}^T$ and ${\bf X}^H$ denote the transpose and
conjugate transpose of matrix $\bf X$, respectively. $\text{Tr}(\bf X)$ denotes the trace of matrix ${\bf X}$. $\text{vec}({\bf X})$ indicates stacking all columns of matrix ${\bf X}$ into a column vector. $\mathbb{E}(X)$ represents the mathematical expectation of
 random variable $X$. $\mathbb{I}_{\{\cdot\}}$ denotes the indicator function. 
 $\log(x)$, 
 and $|x|$ denote the base-2 logarithm, 
 and the norm of $x$, respectively. $\parallel {\bf X} \parallel _F$ represents the Frobenius norm of ${\bf X}$. ${\bf I}_{N}$ denotes the $N$-dimensional identity matrix. $\min\{x, y\}$ indicates the minimum value between two real numbers $x$ and $y$. $j=\sqrt{-1}$ denotes the imaginary unit. $\mathrm{Re}(x)$ denotes the real part of complex number $x$.

\section{System Model}
\begin{figure}[t]
	\setlength{\abovecaptionskip}{0.5cm}
	\centering
	\includegraphics[width=2.4in]{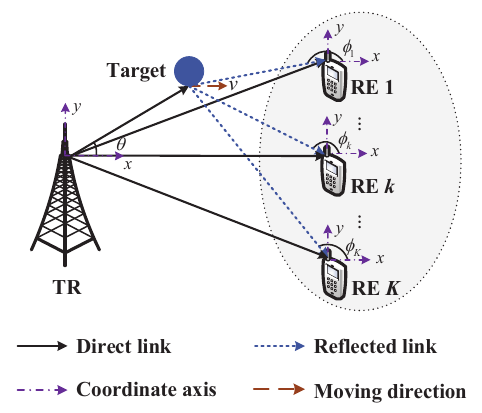}
	\caption{Multi-static ISAC system with RE cooperations.}
	\label{s}
\end{figure}

We consider a multi-static ISAC system with one TR, one target, and $K$ REs, which are deployed in an area of interest, as shown in Fig. \ref{s}. 
The TR sends the ISAC signal to the target and the REs. Each RE receives the reflected signal from the target for sensing (localization) and the signal directly from the TR for communications. 
The TR and the REs are equipped with $N_t$ and $N_r$ antennas, respectively. The target moves at velocity $v$ within the surveillance area. For simplicity, this paper focuses on a simple two-dimensional (2D) scenario to model the considered ISAC system, and the more practical three-dimensional (3D) case can be analyzed in a similar way.
%
%


%

Due to the multi-antenna deployment at the TR and the REs, transmissions of multiple data streams are considered in this paper. 
Let ${\bf s}_k\in\mathbb{C}^{L\times 1}$ denote the communication data vector for the $k$-th RE, $ k\in\mathcal{K}=\{1,\cdots,K\}$, and ${\bf s}_0\in\mathbb{C}^{L\times 1}$ be the dedicated localization data vector for the target, with $L\leq\min\{N_t,N_r\}$ being the number of data streams. 
 Based on the above setup, the transmit signal at the TR in the $t$-th time slot, $t\in \mathcal{T}=\{1,2,\cdots, T\}$, is composed of all the signals for the REs and the target, i.e., 
\begin{align}\label{equ:A}
	{\bf x}(t)&=\sum_{k\in\{0\}\cup\mathcal{K}}{\bf W}_k(t){\bf s}_k(t),
\end{align}
where ${\bf W}_k\in\mathbb{C}^{N_t\times L}, k\in\mathcal{K},$ and ${\bf W}_0\in\mathbb{C}^{N_t\times L}$ represent the transmit beamforming matrices for the $k$-th RE and the target, respectively; ${\bf s}_k$ and ${\bf s}_0$ are modeled as Gaussian distributed complex symmetric circularly Gaussian (CSCG) signal \cite{10086626}, i.e., ${\bf s}_k$ and ${\bf s}_0\sim\mathcal{CN}(0,{\bf I}_L)$, and they are jointly independent. 
 Then, the transmission power at the TR is calculated as
\begin{align}
	\mathbb{E}\left(\parallel {\bf x}(t)\parallel^2\right)=\sum_{k\in\{0\}\cup\mathcal{K}}{\textrm {Tr}}({\bf W}_k(t){\bf W}_k^H(t))\leq P_T,
	\end{align}
where $P_T$ is the power budget.
In order to improve the localization performance, multiple REs with different positions are expected to exchange their obtained information and then cooperatively localize the target.
%

%
\section{ISAC Problem Formulation}
In this section, we first analyze the localization and communication performances, respectively. Then, a CRB minimization problem is formulated by considering the cooperative cost and communication rate requirements.
\subsection{Localization Performance Analysis}
In the $t$-th time slot, the TR sends signal ${\bf x}(t)$ given in (\ref{equ:A}), which is reflected by the target and then arrived at the $K$ REs. Therefore, the received signal at the $k$-th RE is expressed as  
\begin{align}\label{equ: sensing receive}
{\bf y}_k(t)=&{\bf H}_{b,0,k}e^{j2\pi f_kt}{\bf x}(t-\tau_k)+{\bf c}_k(t)+{\bf z}_k(t),
\end{align}
where ${\bf H}_{b,0,k}\in\mathbb{C}^{N_r\times N_t}$ is
 the channel coefficient matrix of the $k$-th TR-target-RE link; $f_k$ is the Doppler shift at the $k$-th RE caused by the target movement;  $\tau_k$ is the transmission delay of the $k$-th TR-target-RE link;
 ${\bf c}_k(t)$ is the clutter interference in the $t$-th time slot including the reflected signal from clutters and the interference from the $k$-th direct TR-RE link, and is modeled as a 
CSCG noise with mean zero and covariance matrix ${\bf C}_k=\sigma_c^2{\bf I}_{N_r}$ \cite{9724174}; and ${\bf z}_k(t)$ is the received CSCG noise at the $k$-th RE, with ${\bf z}_k\sim\mathcal{CN}(0,\sigma_z^2{\bf I}_{N_r})$. 
In this work, we consider the block fading scenario, where the coefficients remain constants within one transmission block, while may vary across different blocks \cite{9391685, 9682613}. Hence, ${\bf H}_{b,0,k}$ is modeled as ${\bf H}_{b,0,k}=\sqrt{\eta_{b,0,k}}\beta_k{\bf a}_k(\phi_k){\bf a}_b(\theta)^T$\footnote{Since the signal transmissions and receptions are analyzed during one typical transmission block in sequel, we omit $t$ in ${\bf H}_{b,0,k}(t)$ to simplify the notations.}, 
where $\eta_{b,0,k}={d_{b,0,k}^{-\epsilon}}$ represents the large-scale loss, with $d_{b,0,k}$ being the distance from the TR to the target and then to the $k$-th RE and $\epsilon$ being the path loss exponent \cite{9682613}; 
 $\beta_k$ is the reflection coefficient for the $k$-th link; 
 ${\bf a}_b(\theta)$ and ${\bf a}_k(\phi_k)$ respectively denote the transmit and receive steering vectors, i.e., 
%
%
%
\begin{align}\label{angle1}
&{\bf a}_b(\theta)=\left[1,e^{j2\pi \frac{d_b}{\lambda}\sin(\theta)}, \cdots, e^{j2\pi \frac{(N_t-1)d_b}{\lambda}\sin(\theta)}\right]^T,\\
	\label{angle2}&{\bf a}_k(\phi_k)=\left[1,e^{j2\pi \frac{d_k}{\lambda}\sin(\phi_k)}, \cdots, e^{j2\pi \frac{(N_r-1)d_k}{\lambda}\sin(\phi_k)}\right]^T, 
\end{align}
with $\theta$, and $\phi_k$ being the DoD of the transmit signal from TR to target, and the DoA of the received signal from the target to the $k$-th RE, respectively;  $d_b$ and $d_k, k\in\mathcal{K},$ being the distances between any two adjacent antennas at the TR and the $k$-th RE, respectively; and $\lambda$ representing the carrier wavelength.

It is worthy to notice that the clutter ${\bf c}_k(t)$ in (\ref{equ: sensing receive}) can be modeled to be signal-dependent or signal-independent \cite{9724174}. In the signal-dependent case, ${\bf c}_k(t)$ is usually modeled as ${\bf c}_k=\sum_{i=1}^{I}{\bf H}_{b,0,i}e^{j2\pi f_it}{\bf x}(t-\tau_i)+{\bf H}_{b,k}{\bf x}(t)$ with $I$ being the number of clutter, and the corresponding covariance matrix can be computed as ${\bf C}_k=\sum_{k\in\{0\}\cup\mathcal{K}}{\text {Tr}}\left({\bf W}_k{\bf W}_k^H\right)\left(\sum_{i=1}^{I}{\bf H}_{b,0,i}{\bf H}_{b,0,i}^H+{\bf H}_{b,k}{\bf H}_{b,k}^H\right)$, for given ${\bf x}$ in (\ref{equ:A}), where ${\bf H}_{b,k}=\sqrt{\eta_{b,k}}{\bf a}_k(\theta_k){\bf a}_b(\theta_k)^T$ being the channel coefficient matrix of the $k$-th direct TR-RE link, and $\theta_k$ is the DoD (and also DoA) of the direct signal from TR to the $k$-th RE. Then, by utilizing conventional iterative algorithms \cite{9652071}, the covariance matrix of ${\bf c}_k(t)$ can be regarded as a constant for fixed ${\bf W}_k$ in the last round of iteration. For simplicity, we directly consider the signal-independent  case, and focus on the constant covariance matrix ${\bf C}_k=\sigma_c^2{\bf I}_{N_r}$ of clutter ${\bf c}_k(t)$ in this work.

\begin{Remark}\label{remark3.1}
	In the considered multi-static ISAC system, each RE only knows the DoA $\phi_k$ (and $\theta_k$) of its received signal and the distance $d_{b,k}$ from the TR to itself, not the DoD $\theta$ of the transmitted signal from TR to target and the distance $d_{0,k}$ from the target to itself. 
	Therefore, in order to locate the target at each RE, it is necessary to find an efficient way to obtain both the DoD $\theta$ and distance $d_{0,k}$ from the target to each RE by utilizing the available knowledge. 
\end{Remark}


\subsubsection{Matched Filtering (MF)}
For the considered multi-static ISAC system, the localization data ${\bf s}_0$ is known for both the TR and the ISAC REs  \cite{10147248}. Therefore, the $k$-th ISAC RE receives signal ${\bf y}_k(t)$ in (\ref{equ: sensing receive}), and then processes it using a matched filter with a delayed and Doppler-shifted version of ${\bf s}_0$, i.e., ${\bf s}_0^H(t-\tau)e^{-j2\pi ft}$. Hence, the output of the matched filter is given as 
 \begin{align}\label{equ:est}
	\hat{\bf \Phi}_k(f,\tau)&=\int_0^{\Delta T}{\bf y}_k(t){\bf s}_0^H(t-\tau)e^{-j2\pi ft}dt\\
	\label{equ:est1}&=\int_0^{\Delta T}{\bf H}_{b,0,k}{\bf W}_0{\bf s}_0(t-\tau_k){\bf s}_0^H(t-\tau)e^{j2\pi (f_k-f)t}dt\nonumber\\
&~~~~+\hat {\bf C}_s+\hat {\bf C}_z,
\end{align}
where $\Delta T$ represents the length of one time slot; and
(\ref{equ:est1}) is obtained by substituting (\ref{equ:A}) and (\ref{equ: sensing receive}) into (\ref{equ:est}). Here, the interference terms ${\bf H}_{b,0,k}\sum_{m\in\mathcal{K}}{\bf W}_{m}{\bf s}_{m}(t-\tau_k){\bf s}_0^H(t-\tau)$ and $({\bf c}_k(t)+{\bf z}_k(t)){\bf s}_0^H(t-\tau)$ cannot be ignored, which is due to ${\bf s}_{m}{\bf s}_0^H, {\bf c}_{k}{\bf s}_0^H, {\bf z}_{k}{\bf s}_0^H \neq {\bf 0}$ with $L$ and $N_r$ being not going to infinity, $\forall k\in\mathcal{K}$ \cite{massive_mimo_proof}. 
Hence, both of them are modeled as noises, i.e., 
$\hat {\bf C}_s = \int_0^{\Delta T}{\bf H}_{b,0,k}\sum_{m\in\mathcal{K}}{\bf W}_{m}{\bf s}_{m}(t-\tau_k){\bf s}_0^H(t-\tau)e^{j2\pi (f_k-f)t}dt$ and $\hat {\bf C}_z = \int_0^{\Delta T}\Big({\bf c}_k(t)+{\bf z}_k(t)\Big){\bf s}_0^H(t-\tau)e^{-j2\pi ft}dt$
 are with mean zero and variances $\hat\sigma_s^2$ and $\hat{\sigma}_z^2$, respectively \cite{haochengh_3D}.
 
%

\begin{figure*}[b]
\hrulefill
\vspace{0in}
 \setcounter{equation}{7}
\begin{align}\label{estimations}
	\left\{\hat f_k, \hat\tau_k\right\}=&\arg\max_{\{f,\tau\}}\left(\frac{\parallel \int_0^{\Delta T}{\bf H}_{b,0,k}{\bf W}_0{\bf s}_0(t-\tau_k){\bf s}_0^H(t-\tau)e^{j2\pi (f_k-f)t}dt\parallel _F^2}{\parallel \hat {\bf C}_s +\hat {\bf C}_z \parallel _F^2}\right).  
	\end{align}
	\begin{align}\label{equ:pdf}\setcounter{equation}{10}
	f({\bf y}_k|{{\boldsymbol\varrho}_k})=\frac{1}{\pi^{M\times N_r}\det({\bf C}_k)}\exp\left[-({\bf y}_k-\boldsymbol\mu_k({\boldsymbol\varrho}_k))^H{\bf C}_k^{-1}({\bf y}_k-\boldsymbol\mu_k({\boldsymbol\varrho}_k))\right]. 
\end{align}
\end{figure*}

From (\ref{equ:est})-(\ref{equ:est1}), the estimation values of Doppler shift $f_k$ and transmission delay $\tau_k$, denoted as $\hat f_k$ and $\hat\tau_k$, are obtained by 
 maximizing the signal-interference-noise ratio of the output signal given in (\ref{equ:est1}), i.e., (\ref{estimations}) shown in the bottom of this page.
Based on (\ref{estimations}), the optimal $\hat f_k$ and $\hat\tau_k$ can be obtained by numerical methods \cite{9947033}. Here, multiple REs are allows to exchange their obtained transmission delay and Doppler shift information about the received signal to cooperatively localize the target.\footnote{In practice, information exchange among the multiple cooperative REs can be realized via a central controller.}
\begin{Remark}
	 In general, time synchronization between the TR and the REs is essential in the multi-static ISAC system, since it significantly affects the localization accuracy \cite{yuanmingt}. This problem can be addressed by utilizing some global synchronization protocols, e.g., Timing-Sync and reference-broadcast systems \cite{Time_Synchronization}, which exploit the broadcast nature of wireless signals to achieve the global time synchronization with high accuracy, and then the TR and the REs can correctly add the timestamp to complete the localization tasks. In this paper, we consider the case that the TR and the REs are perfectly synchronized, and focus more on the RE selection and transmit beamforming  designs.
\end{Remark}
\subsubsection{CRB for Localization}
 As the estimations of $f_k$ and $\tau_k$ are given in (\ref{estimations}), we utilize these information to localize the target (the details will be discussed in section \ref{section5}). Here, we adopt the CRB, a theoretical limit on the variance of any unbiased estimation \cite{yuanmingt}, as the localization performance metric for the joint estimations of $f_k$ and $\tau_k$. 
 Then, we discretize the continuous received signal ${\bf y}_k(t)$ in (\ref{equ: sensing receive}) at each transmission block with the sample duration $\Delta t$, and obtain $M$ independent observations $\left[{\tilde{\bf y}}_{k,1}, \cdots, {\tilde{\bf y}}_{k,M}\right]$, with $M$ being a sufficiently large integer \cite{Estimation_Theory}. Hence, the sampled version of received signal ${\bf y}_k(t)$ in (\ref{equ: sensing receive}) is written as
\begin{align}\label{equ:y_k_dis}\setcounter{equation}{8}
	{{\bf y}}_k=\begin{bmatrix}
		{\tilde{\bf y}}_{k,1}\\
		{\tilde{\bf y}}_{k,2}\\
		\vdots\\
		{\tilde{\bf y}}_{k,M}
	\end{bmatrix} = \begin{bmatrix}
		\boldsymbol\mu_{k,1}({\boldsymbol\varrho}_k)\\
		\boldsymbol\mu_{k,2}({\boldsymbol\varrho}_k)\\
		\vdots\\
		\boldsymbol\mu_{k,M}({\boldsymbol\varrho}_k)
	\end{bmatrix}+\begin{bmatrix}
		{\tilde{\bf c}}_{k,1}\\
		{\tilde{\bf c}}_{k,2}\\
		\vdots\\
		{\tilde{\bf c}}_{k,M}
	\end{bmatrix}+\begin{bmatrix}
		{\tilde{\bf z}}_{k,1}\\
		{\tilde{\bf z}}_{k,2}\\
		\vdots\\
		{\tilde{\bf z}}_{k,M}
	\end{bmatrix},
\end{align}
where ${\bf y}_k$ is the sample version of continuous signal ${\bf y}_k(t)$; ${\tilde{\bf y}}_{k,m}$ is the $m$-th sample of continuous signal ${{\bf y}}_{k}(t), m\in\mathcal{M}=\{1,\cdots,M\}$; $\boldsymbol\mu_{k,m}({\boldsymbol\varrho}_k)={\bf H}_{b,0,k}e^{j2\pi {\tilde f}_km}{\bf x}[m-{\tilde\tau}_k]$, with ${\bf x}[m]=\sum_{k\in\{0\}\cup\mathcal{K}}\sum_{m\in\mathcal{M}}$${\bf W}_k{\bf s}_{k}[m]g(t-(m-1)\Delta t)$ being the $m$-th sample of continuous signal ${\bf x}(t)$ in (\ref{equ:A}), ${\bf s}_k[m]$ being the $m$-th sample of ${\bf s}_k(t)$,
$g(t)$ being the transmit pulse function with $g(t)\in[0,\Delta t]$ and $\frac{1}{\Delta t}{\int_0^{\Delta t}|{g}(t)|^2dt}=1$, $\tilde f_k=f_k\Delta t$, and $\tilde\tau_k=\tau_k/\Delta t$;  ${\tilde{\bf c}}_{k,m}$ and ${\tilde{\bf z}}_{k,m}$ are the $m$-th samples of the continuous clutter ${\bf c}_k(t)$ and noise ${\bf z}_k(t)$, respectively, i.e.,  ${\bf c}_k=\left[{\tilde{\bf c}}_{k,1}, \cdots, {\tilde{\bf c}}_{k,M}\right]$ and ${\bf z}_k=\left[{\tilde{\bf z}}_{k,1}, \cdots, {\tilde{\bf z}}_{k,M}\right]$.


Based on (\ref{equ:y_k_dis}),  the Fisher Information matrix (FIM) ${\bf J}_k({\boldsymbol\varrho}_k)$, which is defined as the inverse of the CRB matrix, is given as \cite{pdf}
%
%
 \begin{align}\label{equ:JJ}
 	{\bf J}_k({\boldsymbol\varrho}_k)=\mathbb{E}_{{\bf y}_k|{\boldsymbol\varrho}_k}\left\{
 	\begin{array}{l}
\left(\frac{\partial}{\partial{{\boldsymbol\varrho}_k}}\ln f({\bf y}_k|{{\boldsymbol\varrho}_k})\right)\\\left(\frac{\partial}{\partial{{\boldsymbol\varrho}_k}}\ln f({\bf y}_k|{{\boldsymbol\varrho}_k})\right)^T
 	\end{array}\right\}.
 \end{align}
Here, $f({\bf y}_k|{{\boldsymbol\varrho}_k})$ is the conditional probability density function of ${\bf y}_k$
expressed as (\ref{equ:pdf}) \cite{pdf},
where $\boldsymbol\mu_k({\boldsymbol\varrho}_k)=[\boldsymbol\mu_{k,1}({\boldsymbol\varrho}_k),\cdots,\boldsymbol\mu_{k,M}({\boldsymbol\varrho}_k)]$; and ${\bf C}_k=(\sigma_c^2+\sigma_z^2){\bf I}_{N_r}$ is the interference covariance matrix, resulting from the clutter ${\bf c}_k$ and noise ${\bf z}_k$ defined in (\ref{equ: sensing receive}).
%
Then, the FIM with respect to ${\boldsymbol\varrho_k}$ defined in (\ref{equ:JJ}) is rewritten as \cite{Estimation_Theory}
\begin{align}\label{equ:J_2}\setcounter{equation}{11}
	{\bf J}_k({\boldsymbol\varrho}_k)=\begin{bmatrix}
{D}_{\tilde \tau_k,\tilde \tau_k} & {D}_{\tilde \tau_k,\tilde f_k}\\
{D}_{\tilde f_k,\tilde\tau_k} & {D}_{\tilde f_k,\tilde f_k}
\end{bmatrix},
\end{align}	
where 
the elements of matrix ${\bf J}({\boldsymbol\varrho}_k)$ in (\ref{equ:J_2}) are defined as
 ${D}_{\tilde\tau_k,\tilde\tau_k}=-\mathbb{E}\left[\frac{\partial^2\ln f({\bf y}_k|{{\boldsymbol\varrho}_k})}{\partial{\tilde\tau_k}^2}\right]$,$ {D}_{\tilde\tau_k,\tilde f_k}=-\mathbb{E}\left[\frac{\partial^2\ln f({\bf y}_k|{{\boldsymbol\varrho}_k})}{\partial{\tilde f_k}\partial{\tilde \tau_k}}\right]$, ${D}_{\tilde f_k,\tilde\tau_k}=-\mathbb{E}\left[\frac{\partial^2\ln f({\bf y}_k|{{\boldsymbol\varrho}_k})}{\partial{\tilde\tau_k}\partial{\tilde f_k}}\right]$, and ${D}_{\tilde f_k,\tilde f_k}=-\mathbb{E}\left[\frac{\partial^2\ln f({\bf y}_k|{{\boldsymbol\varrho}_k})}{\partial{\tilde f_k}^2}\right]$, respectively. 

 \begin{Proposition}\label{pro2}
By taking the second-order partial derivatives of elements in (\ref{equ:J_2}) with respect to $\tilde \tau_k$ and $\tilde f_k$, the closed-form expressions of ${D}_{\tilde\tau_k,\tilde\tau_k}$, ${D}_{\tilde f_k,\tilde f_k}$,  ${D}_{\tilde \tau_k,\tilde f_k}$ and ${D}_{\tilde f_k,\tilde \tau_k}$ in (\ref{equ:J_2}) are respectively derived as 
 \begin{align}
 	\label{equ:D_tt}{D}_{\tilde\tau_k,\tilde\tau_k}&=-\mathbb{E}\left[\frac{\partial^2\ln f({\bf y}_k|{{\boldsymbol\varrho}_k})}{\partial{\tilde\tau_k}^2}\right]=\iota_k\Upsilon_k({\bf W}),\\
 	\label{equ:D_ff}{D}_{\tilde f_k,\tilde f_k}&=-\mathbb{E}\left[\frac{\partial^2\ln f({\bf y}_k|{{\boldsymbol\varrho}_k})}{\partial{\tilde f_k}^2}\right]=\chi_k\Upsilon_k({\bf W}),\\
	\label{equ:D_tf1}{D}_{\tilde \tau_k,\tilde f_k}&=-\mathbb{E}\left[\frac{\partial^2\ln f({\bf y}_k|{{\boldsymbol\varrho}_k})}{\partial{\tilde \tau_k}{\partial \tilde f_k}}\right]
	=\varsigma_k\Upsilon_k({\bf W}), \\
	\label{equ:D_tf}{D}_{\tilde f_k, \tilde \tau_k}&=-\mathbb{E}\left[\frac{\partial^2\ln f({\bf y}_k|{{\boldsymbol\varrho}_k})}{\partial{\tilde f_k}{\partial \tilde \tau_k}}\right]=\varsigma_k\Upsilon_k({\bf W}),
\end{align}
where $\Upsilon_k({\bf W})$ is expressed by $\Upsilon_k({\bf W})=
	 \textrm{Tr}\left({\bf H}_{b,0,k}^{H}{\bf H}_{b,0,k}\sum_{k\in\{0\}\cup\mathcal{K}}{\bf W}_k{\bf W}_k^H\right)$, with ${\bf W}=\{{\bf W}_0,\cdots, {\bf W}_K\}$ being the beamforming matrix for the target and the $K$ REs; $\iota_k =\frac{4BMF_g\eta_k}{(1+\alpha_k)(\sigma_c^2+\sigma_z^2)}$, $\chi_k=\frac{64\pi^2B^3MF_{tg}\eta_k}{(1+\alpha_k)(\sigma_c^2+\sigma_z^2)}$, and $\varsigma_k= \frac{16\pi B^2MF_{t\dot g}\eta_k}{(1+\alpha_k)(\sigma_c^2+\sigma_z^2)}$, with
$B$ being the bandwidth of transmitted signal; and $F_g, F_{tg}$, and $F_{t\dot g}$ are computed as $F_g={\int_0^{\Delta t}|\dot{g}(t)|^2dt}$, 
$F_{tg}=\int_{0}^{\Delta t}t^2|g(t)|^2dt$, $F_{t\dot g}=\int_{0}^{\Delta t}tg(t){\dot g}^*(t)dt$, respectively, with ${\dot g}=\frac{\partial g(t)}{\partial t}$, ${\dot g^*}=\frac{\partial g^*(t)}{\partial t}$, and $g(t)$ being defined in (\ref{equ:y_k_dis}).
%
%
%
%
 \end{Proposition}
 \begin{IEEEproof}
Please see Appendix A.	
\end{IEEEproof}
Based on the above analysis, CRB ${U}({\bf b},{\bf W})$, which is defined as trace of the inverse of FIM \cite{10146016}, is adopted as the localization performance metric for the considered multi-static ISAC system, i.e., (\ref{equ:crb_u}),
where 
${\bf b}= [b_1,\cdots,b_K]^T$ is a binary vector, i.e., $b_k=1$ indicates the $k$-th RE being selected for localization and otherwise, we have $b_k=0$; ${\bf J}_k({\boldsymbol\varrho}_k)$ is given in (\ref{equ:J_2}); 
and (\ref{equ:crb_u2}) is obtained by substituting ${D}_{\tilde\tau_k,\tilde\tau_k}$, ${D}_{\tilde f_k,\tilde f_k}$,  ${D}_{\tilde \tau_k,\tilde f_k}$, and ${D}_{\tilde f_k,\tilde \tau_k}$ in (\ref{equ:D_tt})-(\ref{equ:D_tf}) into (\ref{equ:crb_u}).

\begin{figure*}[b]
\hrulefill
\vspace{0in}
 \setcounter{equation}{16}
\begin{align}\label{equ:crb_u}
	{U}({\bf b},{\bf W})&={\textrm {Tr}}\left\{\Big(\sum_{k\in\mathcal{K}}b_k{\bf J}_k({\boldsymbol\varrho}_k)\Big)^{-1}\right\}\\
	\label{equ:crb_u2}&=\frac{\sum_{k\in\mathcal{K}}b_k(\chi_k+\iota_k)\Upsilon_k({\bf W})}{\left(\sum_{k\in\mathcal{K}}b_k\chi_k\Upsilon_k({\bf W})\right)\left(\sum_{k\in\mathcal{K}}b_k\iota_k\Upsilon_k({\bf W})\right)-\left(\sum_{k\in\mathcal{K}}b_k\varsigma_k\Upsilon_k({\bf W})\right)^2}.
\end{align} 
 \setcounter{equation}{20}
\begin{align}\label{equ:Psi}
	{\bf \Psi}_k=\begin{cases}
		\sum_{k^{\prime}\in\mathcal{K}\backslash \{k\}}{\bf H}_{b,k}{\bf W}_{k^{\prime}}{\bf W}_{k^{\prime}}^H{\bf H}_{b,k}^H
	+\sigma^2{\bf I}_{N_r} & b_k=1,\\
	\sum_{k^{\prime}\in\mathcal{K}\backslash \{k\}}{\bf H}_{b,k}{\bf W}_{k^{\prime}}{\bf W}_{k^{\prime}}^H{\bf H}_{b,k}^H
	+{\bf H}_{b,k}{\bf W}_{0}{\bf W}_{0}^H{\bf H}_{b,k}^H+\sigma^2{\bf I}_{N_r}, & b_k=0.
	\end{cases}
\end{align}
\end{figure*}

\subsection{Communication Performance Analysis}
The communications from the TR
to each RE can be modeled as a multi-input multi-output channel \cite{10146016}. Then, the received signal ${\bf y}_k^{c}$ at the $k$-th RE includes the desired signal from the TR, the localization interference signal, the communication interference from the TR, and the CSCG noise, i.e., 
 \setcounter{equation}{18}
\begin{align}\label{transmit_signal}
	{\bf y}_k^{c}&={\bf H}_{b,k}{\bf x}+{\bf n}_k,\nonumber \\
&=\underbrace{{\bf H}_{b,k}{\bf W}_k{\bf s}_k}_{\text{Desired  signal}}+\underbrace{{\bf H}_{b,k}{\bf W}_0{\bf s}_0}_{\text{Localization interference}}+\underbrace{{\bf H}_{b,k}\sum_{k^{\prime}\in\mathcal{K}\backslash \{k\}}{\bf W}_{k^{\prime}}{\bf s}_{k^{\prime}}}_{\text{communication interference}}\nonumber\\
&~~~~+{\bf n}_k,
\end{align}
where ${\bf x}$ is the transmit signal ${\bf x}(t)$ in (\ref{equ:A}) by omitting index $t$; ${\bf H}_{b,k}\in\mathbb{C}^{N_r\times N_t}$ is the channel coefficient matrix from the TR to the $k$-th RE;
 the first term in (\ref{transmit_signal}) is the desired signal of the $k$-th RE; the second term in (\ref{transmit_signal}) is the localization interference and can be perfectly canceled if the $k$-th RE is selected for localization, and otherwise, it is treated as noise \cite{10086626}; the third term in (\ref{transmit_signal}) is the communication interference; and the fourth term ${\bf n}_k$ in (\ref{transmit_signal}) is the CSCG noise with ${\bf n}_k\sim\mathcal{CN}(0,\sigma^2{\bf I}_{N_r})$.


Based on the above analysis, the transmission rate for the  communications at the $k$-th RE is calculated as 
\begin{align}\label{equ:transmission_rate}
	R_k({\bf b},{\bf W})=\log\det\left({\bf I}_{L}+{\bf W}_{k}^H{\bf H}_{b,k}^H{\bf \Psi}_k^{-1}{\bf H}_{b,k}{\bf W}_{k}\right),  
\end{align}
with ${\bf \Psi}_k$ being given in (\ref{equ:Psi}).
Here, if $b_k=1$, the $k$-th RE is selected for localization and otherwise, it only works as a communication RE.
\begin{Remark}
	From (\ref{equ:transmission_rate})-(\ref{equ:Psi}),
	it is observed that when the $k$-th RE is selected for localization, there is no localization interference, resulting in higher achievable communication rate. From (\ref{equ:crb_u2}), it is also revealed that the CRB decreases as the number of the selected REs increases. However, multi-RE cooperatively localization also causes high cooperation cost. Hence, it is necessary to investigate the localization performance with limited cooperation cost and minimum communication rate requirement.
	\end{Remark}

\subsection{Problem Formulation}
Without loss of generality, the cooperation cost in the considered multi-static ISAC system is modeled as the sum  of the prices of all selected REs \cite{6031934,10146016},
i.e., 
 \setcounter{equation}{21}
\begin{align}
	\Omega({\bf b})=\sum_{k\in\mathcal{K}}{\upsilon}_k{b}_k,
\end{align}
where $\upsilon_k>0$ is the cooperation price by involving the $k$-th RE for cooperative localization. Intuitively, the RE close to the target and deployed near other REs is more likely to be selected for localization and is thus assigned with a lower price \cite{6031934}. Hence, we model $\upsilon_k$ as the weighted sum of distance from itself to the target and the average distance from itself to other cooperative REs, i.e., $\upsilon_k=\rho{d_{0,k}}+(1-\rho)\frac{\sum_{k\neq k^{\prime},k,k^{\prime}\in\mathcal{G}}d_{k,k^{\prime}}}{|\mathcal{G}|-1}$, with $\rho\in[0,1]$ being a weight factor, and $\mathcal{G}=\{k\in\mathcal{K}\vert b_k=1\}$ being the set of all selected REs for localization. 

%
%
%

Our goal is to minimize the CRB for localization and guarantee the  communication rates by jointly design the binary vector ${\bf b}$ for RE selection and the transmit beamforming matrix ${\bf W}$ at the TR. Then, the CRB minimization problem is formulated as 
\begin{align}\label{equ:D3}
	\min_{\{\bf b, W\}} ~~~~&~{U}({\bf b},{\bf W})\\
	\label{equ:D3-2}\textrm{s.t}.~~~~~&~R_k({\bf b},{\bf W})\geq R_{th}, ~~\forall k\in\mathcal{K},\\
	\label{equ:D3-1}~~~~&~\sum_{k\in\{0\}\cup\mathcal{K}}{\textrm {Tr}}({\bf W}_k{\bf W}_k^H)\leq P_T,\\
	 \label{equ:D3-4}~~~~&~\Omega({\bf b})\leq \Omega_{th},\\
	  \label{equ:D3-3}&~b_k\in\{0,1\}, ~~\forall k\in\mathcal{K},
\end{align}
where (\ref{equ:D3-2}) is the communication rate constraint for each RE, with $R_{th}$ being the minimum communication rate requirement; (\ref{equ:D3-1}) is the power constraint at the TR, with $P_T$ being the power budget; (\ref{equ:D3-4}) is the cooperation cost constraint, with $\Omega_{th}$ denoting the maximum cost; and (\ref{equ:D3-3}) is the constraint for RE selection. 
It is obvious that objective function in (\ref{equ:D3}) and  constraint in (\ref{equ:D3-2}) are non-convex, making problem (\ref{equ:D3})-(\ref{equ:D3-3}) non-convex and  difficult to be solved \cite{10287247}.
%
%
%
%
Moreover, binary variable $b_k, k\in\mathcal{K}$, also makes problem (\ref{equ:D3})-(\ref{equ:D3-3}) a mixed-integer program \cite{10173720}, which is  intractable in general. Therefore, it is necessary to design an efficient method to find a near-optimal solution for problem (\ref{equ:D3})-(\ref{equ:D3-3}). 
%
%

%
%
%
%

\section{Algorithms}
 To address the above challenges, problem (\ref{equ:D3})-(\ref{equ:D3-3}) is decoupled as a RE selection subproblem, which is solved by a minimax linkage-based method, and a transmit beamforming optimization subproblem with non-convex constraints, which is solved by a successive convex approximation algorithm.
%
%

\subsection{RE Selection}
For fixed transmit beamforming matrix $\bar{\bf W}$, the original problem (\ref{equ:D3})-(\ref{equ:D3-3}) is simplified into the following RE selection problem, i.e., 
 \setcounter{equation}{28}
\begin{align}\label{equ:D4}
	\min_{\bf b} ~~~~&~{U}({\bf b})\\
	\label{equ:D4-2}\textrm{s.t}.~~~~~&~R_k({\bf b})\geq R_{th}, ~~\forall k\in\mathcal{K},\\
	 \label{equ:D4-4}~~~~&~\text{(\ref{equ:D3-4})},\text{(\ref{equ:D3-3})},
\end{align}
where ${U}({\bf b})$ and $R_k({\bf b})$ are obtained by substituting a fixed $\bar{\bf W}$ into (\ref{equ:crb_u2}) and (\ref{equ:transmission_rate}), respectively. Here,
exhaustive search can be employed to solve problem (\ref{equ:D4})-(\ref{equ:D4-4}), aiming to find the optimal solution for all REs satisfying the constraints in (\ref{equ:D4-2})-(\ref{equ:D4-4}). However, it should be noted that the computational complexity of this method is $\mathcal{O}(2^K)$, which grows exponentially with the number of REs $K$. To address this challenge, a minimax linkage-based RE selection method is proposed in this paper to solve problem (\ref{equ:D4})-(\ref{equ:D4-4}) with a much lower computational complexity of $\mathcal{O}(K^3)$ \cite{9381616}.

\begin{algorithm}[h] 
		\caption{ Minimax linkage-based RE selection for problem (\ref{equ:D4})-(\ref{equ:D4-4}).\label{algorithm1}}
		\vspace{0.1cm}
		\begin{algorithmic}[1]
			\State Initialize the set of point $\mathcal{P}=\{p_1,\cdots,p_K\}$;
			\State Initialize $S=\{\{p_1\},\cdots,\{p_K\}\}$, and $\mathcal{G}_k=\{p_k\}, \forall k\in\mathcal{K}$; 
			\State Initialize beamforming matrix $\bar{\bf W}$, minimum communication rate requirement $R_{th}$, maximum cooperation cost $\Omega_{th}$, $j=K$, and $\mathcal{U}_0=\emptyset$;
			\State Compute $d(\{p_k\},\{p_{k^{\prime}}\})=d(p_k,p_{k^{\prime}})$ according to Definition \ref{def1}, $\forall p_k, p_{k^{\prime}}\in\mathcal{P}$;
			\State \textbf {While} {$\mathcal{G}_j\neq \mathcal{P}$} \textbf{do}
			\State \quad \ \ Update $j\leftarrow j+1$;
			     \State \quad \ \ Compute $(\mathcal{P}_1,\mathcal{P}_2)=\arg\min_{X,Y\in S,X\neq Y}d(X,Y)$ according to Definition \ref{def4};
			     \State \quad \ \ Set $\mathcal{G}_j=\{\mathcal{P}_1\cup \mathcal{P}_2\}$;
			     \State \quad \ \ Update $S\leftarrow \{S,\{\mathcal{P}_1\cup \mathcal{P}_2\}\}\backslash \{\mathcal{P}_1,\mathcal{P}_2\}$;  
			     \State \quad \ \ Compute $d(\mathcal{P}_1\cup\mathcal{P}_2, X), \forall X \in S$;
			\State \textbf {End while}
			\State Output  groups $\{\mathcal{G}_1,\cdots,\mathcal{G}_K,\mathcal{G}_{K+1},\cdots\mathcal{G}_{j}\}$;
			\State \textbf {For} {$n=1,\cdots,j$} \textbf{do}
			\State \quad \ \ Set ${\bf b}_n = {\bf 0}$; 
			\State \quad \ \ Update ${\bf b}_n$ by setting $b_k=1$, $\forall p_k \in \mathcal{G}_n$; 
			 \State \quad \ \ Compute $U({\bf b}_n), R_k({\bf b}_n)$ and $\Omega({\bf b}_n)$ according to (\ref{equ:D4})-(\ref{equ:D4-4});
			\State \quad \ \ \textbf {If} {(\ref{equ:D4-2}) and (\ref{equ:D4-4}) hold} \textbf{then}
			\State \quad \quad \ \ \ \ Update $\mathcal{U}_0\leftarrow \mathcal{U}_0\cup \{U({\bf b}_n)\}$;
			\State \quad \ \ \textbf {End if}
			\State \textbf {End for}
			\State Obtain RE selection strategy ${\bf b}=\arg\min_{{\bf b}_n} \mathcal{U}_0$.
		\end{algorithmic}
\end{algorithm}

Unsimilar to conventional cooperative works that only utilized the distances among cooperative REs to model the cooperation cost \cite{9381616,10168186,9771660}, the proposed minimax linkage-based RE selection method further incorporates the distances from the target to the REs.
%
%
Let $d:\mathbb{R}^2\times\mathbb{R}^2\rightarrow \mathbb{R}$ be the Euclidean distance function, and 
$\mathcal{P}$ be the set of some points in $\mathbb{R}^2$, i.e., the locations of all REs, and we then define the following concepts.
\begin{Definition} \label{def1}
The maximal distance between the $k$-th RE and the set $\mathcal{P}$ is defined as the farthest point in $\mathcal{P}$ to the $k$-th RE, i.e., $d_{max}(p_k,\mathcal{P})=\max_{p_{k^{\prime}}\in\mathcal{P}}d(p_k,p_{k^{\prime}})$, with $p_k=(x_k,y_k)$ being the position of the $k$-th RE, and $d(p_k,p_{k^{\prime}})=\sqrt{(x_k-x_{k^{\prime}})^2+(y_k-y_{k^{\prime}})^2}$.
\end{Definition}
\begin{Definition}
	The minimax distance of set $\mathcal{P}$ is defined as $r_{min}(\mathcal{P})=\min_{p_k\in\mathcal{P}}d_{max}(p_k,\mathcal{P})$.
\end{Definition}
\begin{Definition}
	The joint minimax and nearest distance from the target to set $\mathcal{P}$ is defined as $r(\mathcal{P}) = (1-\rho) r_{min}(\mathcal{P})+\rho\min_{p_k\in\mathcal{P}}d(p_k,p_0)$, with $p_0$ being the location of target, and $\rho\in[0,1]$ being a coefficient to balance the distances from the $k$-th RE to the other cooperative REs and to the target.
\end{Definition}
\begin{Definition}\label{def4}
	The minimax linkage between two sets $\mathcal{P}_1$ and $\mathcal{P}_2$ in $\mathbb{R}^2$ is defined as $d(\mathcal{P}_1,\mathcal{P}_2) = r(\mathcal{P}_1\bigcup \mathcal{P}_2)$.
\end{Definition}


%
The proposed minimax linkage-based RE selection method is summarized in Algorithm \ref{algorithm1}. We set $K$ REs as the initial $K$ groups, i.e., $\mathcal{G}_1=\{p_1\},\dots,\mathcal{G}_K=\{p_K\}$. Then, the linkage criterion merges any two groups that have the smallest joint minimax and nearest distance out of all merging possibilities step by step, and finally generates a linkage tree to connect $K$ REs. 
Here, the $K$ leaf nodes of the tree correspond to the $K$ initial groups, and the non-leaf nodes correspond to the merged groups denoted as $\mathcal{G}_{K+1},\cdots,\mathcal{G}_{j}$, with $j$ being the number of all groups. From the $j$ candidates, we choose the RE group, which has lowest CRB and satisfies constraints (\ref{equ:D4-2}) and (\ref{equ:D4-4}), to cooperatively localize the target.

 \begin{figure*}[b]
\hrulefill
\vspace{0in}
\setcounter{equation}{33}
\begin{align}\label{equ:re_crb}
	{U}_{\mathcal{G}}({\bf W})&=\frac{\sum_{k\in\mathcal{G}}(\chi_k+\iota_k)\Upsilon_k({\bf W})}{\left(\sum_{k\in\mathcal{G}}\chi_k\Upsilon_k({\bf W})\right)\left(\sum_{k\in\mathcal{G}}\iota_k\Upsilon_k({\bf W})\right)-\left(\sum_{k\in\mathcal{G} }\varsigma_k\Upsilon_k({\bf W})\right)^2}.
\end{align} 
 \setcounter{equation}{34}
\begin{align}\label{equ:max1}
	\max_{\bf W}~~\frac{\left(\sum_{k\in\mathcal{G}}\chi_k\Upsilon_k({\bf W})\right)\left(\sum_{k\in\mathcal{G}}\kappa_1\chi_k\Upsilon_k({\bf W})\right)-\left(\sum_{k\in\mathcal{G}}\kappa_2\chi_k\Upsilon_k({\bf W})\right)^2}{(1+\kappa_1)\sum_{k\in\mathcal{G}}\chi_k\Upsilon_k({\bf W})}. 
\end{align}
 
\end{figure*}
\subsection{Transmit Beamforming}
With fixed RE selection strategy ${\bf b}$, the original problem in (\ref{equ:D3})-(\ref{equ:D3-3}) is simplified as the following optimal transmit beamforming problem
 \setcounter{equation}{31}
\begin{align}\label{equ:D5}
	\min_{\bf W} ~~~~&~{U}_{\mathcal{G}}({\bf W})\\
	\label{equ:D5-2}\textrm{s.t}.~~~~&~\text{(\ref{equ:D3-1})}, R_k({\bf W})\geq R_{th}, ~~\forall k\in\mathcal{K},
\end{align}
where ${U}_{\mathcal{G}}({\bf W})$ and $R_k({\bf W})$ are obtained by substituting a fixed $\bar{\bf b}$ into (\ref{equ:crb_u2}) and (\ref{equ:transmission_rate}), respectively. Note that problem (\ref{equ:D5})-(\ref{equ:D5-2}) is still non-convex due to the non-convex objective function in (\ref{equ:D5}) and the communication rate constraint in (\ref{equ:D5-2}). To tackle this problem, we first equivalently transform the objective function into a convex form, and then approximate the communication rate constraint in (\ref{equ:D5-2}) as a convex one.

\subsubsection{Equivalent transformation of objective function} For fixed $\bar{\bf b}$, (\ref{equ:D5}) is rewritten as (\ref{equ:re_crb}),
where $\chi_k, \iota_k, \varsigma_k,$ and ${\Upsilon}_k({\bf W}), \forall k\in \mathcal{G}$, are defined in Proposition \ref{pro2}. It is obvious that (\ref{equ:re_crb}) is non-convex and thus difficult to be solved. 
Fortunately, since the CRB is the minimum variance of any unbiased estimation, i.e., ${U}_{\mathcal{G}}({\bf W})>0$ always holds, it is observed that minimizing 
(\ref{equ:re_crb}) is equivalent to maximizing $\frac{1}{{U}_{\mathcal{G}}({\bf W})}$, i.e., (\ref{equ:max1}),
where $\kappa_1=\frac{\iota_k}{\chi_k}=\frac{F_g}{16\pi^2 B^2F_{tg}}, \kappa_2=\frac{\varsigma_k}{\chi_k}=\frac{F_{t\dot g}}{4\pi BF_{tg}}, \forall k\in\mathcal{K}$, with $F_g={\int_0^{\Delta t}|\dot{g}(t)|^2dt}, F_{tg}=\int_{0}^{\Delta t}t^2|g(t)|^2dt$, and $F_{t\dot g}=\int_{0}^{\Delta t}tg(t){\dot g}^*(t)dt$ being given in Proposition \ref{pro2}.
Since $\chi_k>0$ and $\Upsilon_k({\bf W})> 0$ always hold (as defined in Proposition \ref{pro2}), problem (\ref{equ:max1}) is further equivalent to maximizing a convex version with respect to design variable ${\bf W}$, i.e., 
 \setcounter{equation}{35}
\begin{align}\label{equ:maxfinal}
	\max_{\bf W}~~\frac{\kappa_1-\kappa_2^2}{1+\kappa_1}\sum_{k\in\mathcal{G}}\chi_k\Upsilon_k({\bf W}),
\end{align}
where $\chi_k$ is a constant, and $\Upsilon_k({\bf W})$ is a convex function with respect to $\bf W$, as defined in Proposition \ref{pro2}.

Based on the above analysis, problem (\ref{equ:D5})-(\ref{equ:D5-2}) is equivalent to 
\begin{align}\label{equ:D7}
\max_{{\bf W}}~~&\frac{\kappa_1-\kappa_2^2}{1+\kappa_1}\sum_{k\in\mathcal{G}}\chi_k\Upsilon_k({\bf W})~~~~~~~~~\\
	\label{equ:D7-11}\textrm{s.t.}~~~&R_k({\bf W})-R_{th}\geq 0, ~~\forall k\in\mathcal{K},\\
	\label{equ:D7-2}&P_T-\sum_{k\in\mathcal{K}}{\textrm {Tr}}({\bf W}_k{\bf W}_k^H)\geq 0,
\end{align}
where (\ref{equ:D7}) is directly obtained from (\ref{equ:maxfinal}); and
 (\ref{equ:D7-11}) and (\ref{equ:D7-2}) are obtained by rearranging the terms of (\ref{equ:D3-1}) and the communication rate constraint in  (\ref{equ:D5-2}), respectively. 
However, problem (\ref{equ:D7})-(\ref{equ:D7-2}) is still non-convex due to the non-convexity constraint (\ref{equ:D7-11}).

\subsubsection{Approximation of constraint (\ref{equ:D7-11})}

Giving fixed $\bar{\bf b}$, rate function $R_k({\bf W})$ in (\ref{equ:D7-11}) can be rewritten as 
\begin{align}\label{equ:zzz1}
	R_k({\bf W})&=\log\det\left[{\bf\Psi}_k^{-1}({\bf\Psi}_k+{\bf H}_{b,k}{\bf W}_k{\bf W}_k^H{\bf H}_{b,k}^H)\right]\\
	\label{equ:zzz2}&=\log\det\left({\bf\Psi}_k+{\bf H}_{b,k}{\bf W}_k{\bf W}_k^H{\bf H}_{b,k}^H\right)\nonumber\\
	&~~~~-\log\det({\bf\Psi}_k),
\end{align}
where 
(\ref{equ:zzz1}) holds due to $\det({\bf I}+{\bf XY})=\det({\bf I}+{\bf YX})$; and (\ref{equ:zzz2}) holds due to $\det({\bf Y}^{-1}{\bf X})=\frac{\det({\bf X})}{\det({\bf Y})}$. Then, by substituting (\ref{equ:Psi}) into (\ref{equ:zzz2}), $R_k({\bf W})$ is rewritten as
\begin{align}\label{equ:zzz3}
R_k({\bf W})&=\log\det\left(\sigma^2{\bf I}_{N_r}+\sum_{i\in{\mathcal{ K}^{(\cdot)}}}{\bf W}_i^H{\bf H}_{b,k}^H{\bf H}_{b,k}{\bf W}_i\right)\nonumber\\
&~~~~-\log\det({\bf\Psi}_k^{(\cdot)}),
\end{align}
%
%
%
where $\mathcal{K}^{(\cdot)}=\mathcal{K}$ and ${\bf\Psi}_k^{(\cdot)}={\bf\Psi}_k^{\text{\MakeUppercase{\romannumeral 1}}}$ if $b_k=1$, and otherwise we have $\mathcal{K}^{(\cdot)}=\{0\}\cup\mathcal{K}$ and ${\bf\Psi}_k^{(\cdot)}={\bf\Psi}_k^{\text{\MakeUppercase{\romannumeral 2}}}$, with ${\bf\Psi}_k^{\text{\MakeUppercase{\romannumeral 1}}}$ and ${\bf\Psi}_k^{\text{\MakeUppercase{\romannumeral 2}}}$
being obtained by substituting $b_k=1$ and $b_k=0$ into ${\bf \Psi}_k$ given in (\ref{equ:transmission_rate}), respectively.		
%
%
%
As the first term in the right hand of (\ref{equ:zzz3})
is a concave function, we utilize the first-order Taylor expansion to approximate the second term $\log\det({\bf\Psi}_k^{(\cdot)})$, i.e.,
\begin{align}\label{equ:upper_bound}
&\log\det({\bf\Psi}_k^{(\cdot)})\leq\log\det(\bar{\bf\Psi}_k^{(\cdot)})\nonumber\nonumber\\
&+\textrm{Tr}\left((\bar{\bf\Psi}_k^{(\cdot)})^{-1}\sum_{i\in\mathcal{K}^{(\cdot)}}{\bf H}_{b,k}{\bf W}_i{\bf W}_i^H{\bf H}_{b,k}^H\right)\nonumber\\
&-\textrm{Tr}\left((\bar{\bf\Psi}_k^{(\cdot)})^{-1}\sum_{i\in\mathcal{K}^{(\cdot)}}{\bf H}_{b,k}\bar{\bf W}_i\bar{\bf W}_i^H{\bf H}_{b,k}^H\right),
\end{align} 
where both $\bar{\bf\Psi}_k^{\text{\MakeUppercase{\romannumeral 1}}}$ and $\bar{\bf W}_i$ are fixed; and (\ref{equ:upper_bound}) holds due to the fact of $\log\det({\bf I}+{\bf X})\leq \log\det({\bf I}+\bar{\bf X})
+\textrm{Tr}\left[({\bf I}+\bar{\bf X})^{-1}({\bf X}-\bar{\bf X})\right]$. 

By substituting (\ref{equ:upper_bound}) into (\ref{equ:zzz3}), (\ref{equ:D7-11}) is approximated by the following concave function, i.e., 
\begin{align}\label{equ:lower_bound}
&\log\det\left(\sigma^2{\bf I}_{N_r}+\sum_{i\in\mathcal{K}^{(\cdot)}}{\bf H}_{b,i}{\bf W}_i{\bf W}_i^H{\bf H}_{b,i}^H\right)-c_k^{(\cdot)}-R_{th}-\nonumber\\
&\textrm{Tr}\left((\bar{\bf\Psi}_k^{\text{\MakeUppercase{\romannumeral 1}}})^{-1}\sum_{i\in\mathcal{K}^{(\cdot)}}{\bf H}_{b,i}{\bf W}_i{\bf W}_i^H{\bf H}_{b,i}^H\right)\geq 0, \forall k\in\mathcal{K},
\end{align}
where $c_k^{(\cdot)} =\log\det(\bar{\bf\Psi}_k^{\text{\MakeUppercase{\romannumeral 1}}})-\textrm{Tr}\left((\bar{\bf\Psi}_k^{\text{\MakeUppercase{\romannumeral 1}}})^{-1}\sum_{i\in\mathcal{K}}{\bf H}_{b,k}\bar{\bf W}_i\bar{\bf W}_i^H{\bf H}_{b,k}^H\right)$ if $b_k=1$, and otherwise we have $c_k^{(\cdot)} = \log\det(\bar{\bf\Psi}_k^{\text{\MakeUppercase{\romannumeral 2}}})-\textrm{Tr}\left((\bar{\bf\Psi}_k^{\text{\MakeUppercase{\romannumeral 2}}})^{-1}\sum_{i\in\{0\}\cup{\mathcal{K}}}{\bf H}_{b,k}\bar{\bf W}_i\bar{\bf W}_i^H{\bf H}_{b,k}^H\right)$.
\begin{Proposition}
	We replace the non-convex constraints (\ref{equ:D7-11}) in problem (\ref{equ:D7})-(\ref{equ:D7-2}) with its approximative convex version (\ref{equ:lower_bound}), thereby shrinking the feasible region. As a result, we obtain a sub-optimal solution for problem (\ref{equ:D7})-(\ref{equ:D7-2}).
\end{Proposition}

Then, for fixed ${\bar {\bf b}}$, the optimal transmit beamforming sub-problem (\ref{equ:D5})-(\ref{equ:D5-2}) is approximated as 
\begin{align}\label{equ:D6}
	\max_{\bf W} ~~~~&\frac{\kappa_1-\kappa_2^2}{1+\kappa_1}\sum_{k\in\mathcal{G}}\chi_k\Upsilon_k({\bf W})\\
\textrm{s.t.}~~~~
\label{equ:D6-2}&~\text{(\ref{equ:D7-2})},\text{(\ref{equ:lower_bound})}.
\end{align}
It is easy to observed that the objective function and constraints (\ref{equ:D6})-(\ref{equ:D6-2}) are all concave functions with respect to $\bf W$. Hence, problem (\ref{equ:D6})-(\ref{equ:D6-2}) is convex and can be efficiently solved by some standard convex optimization tools, e.g., CVX \cite{8647900}.

\section{Design of Localization Schemes} \label{section5}
Based on the electrodynamic theory, if an RE is moving with velocity $v$ relatively to a TR of frequency $f_0$, the frequency received at the RE is given as \cite{ELECTRODYNAMICS_OF_MOVING_BODIES}
\begin{align}\label{equ:cc}
	f^{\prime}=f_0\frac{1-\zeta\cos(\theta)}{\sqrt{1-\zeta^2}},
\end{align}
where  $\zeta=v/c$ is the ratio of the speed $v$ of the target over that of light in free space denoted as $c$, and $\theta$ is the angle between the $x$ axis and the TR-RE link.


Based on (\ref{equ:cc}), the Doppler shift of each TR-target-RE link in our considered multi-static ISAC system can be regarded as a two-step relativistic Doppler effect.
Therefore, the overall Doppler shift  observed at the $k$-th RE is derived as 
\begin{align}\label{equ:f_k1}
	f_{k}=&f_0\frac{1-\varrho\cos(\theta)}{\sqrt{1-\zeta^2}}\frac{\sqrt{1-\zeta^2}}{1+\zeta\cos(\phi_k)}-f_0\\
	\label{equ:f_k2}=&-\zeta f_0\frac{\cos(\theta)+\cos(\phi_k)}{1+\zeta\cos(\phi_k)}\\
	\label{equ:f_k3}\approx & -\frac{2f_0}{c}\cdot v\cos\left(\frac{\theta+\phi_k}{2}\right)\cdot \cos\left(\frac{\theta-\phi_k}{2}\right),
\end{align}
where $f_0$ is the frequency of transmission signal ${\bf x}$, and $\theta, \phi_k,k\in\mathcal{K},$ are the DoD of the transmit signal at the TR and the  DoA of the received signal at the $k$-th RE, respectively;  (\ref{equ:f_k2}) is obtained by rearranging the terms in (\ref{equ:f_k1}); and (\ref{equ:f_k3}) holds due to the fact of $v\ll c$, i.e., $\zeta=v/c \ll 1$. 
Similarly, the signal transmission delay  received at the $k$-th RE is computed as 
\begin{align}\label{equ:tau_k}
	\tau_k=\frac{d_{b,0}+d_{0,k}}{c},
\end{align}
where $d_{b,0}$ is the distance from the TR to the target, and $d_{0,k}$ is that from the target to the $k$-th RE.
 

\begin{Remark}
	From (\ref{equ:f_k1})-(\ref{equ:tau_k}), it is observed that Doppler shift $f_k$ is a function with respect to $\theta$, and transmission delay $\tau_k$ is a function with respect to  $d_{0,k}$. Hence, we can effectively derive $\theta$ and $d_{0,k}$ by utilizing $f_k$ and $\tau_k$ to localize the target.	
\end{Remark}
\begin{Proposition}\label{pro3}
	Giving the Doppler shifts and the transmission delay received at any two REs, the closed-form expressions of DoD $\theta$ and distance $d_{0,k}$ between the target and the $k$-th RE are respectively calculated as
	\begin{align}\label{11111}
			&\theta =2\wp\left(
			\begin{array}{l}	
\Xi_{k^{\prime}}-\Xi_{k}\cos\left(\frac{\phi_{k}-\phi_{k^{\prime}}}{2}\right),\\\Xi_k\sin\left(\frac{\phi_{k}-\phi_{k^{\prime}}}{2}\right)
\end{array}\right)-\phi_k,\\
		&d_{0,k}=\frac{c^2\tau_k^2-2c\tau_k d_{b,k}\cos(\theta-\wp({x_k-x_b}, {y_k-y_b}))}{2c\tau_k-2d_{b,k}\cos(\theta-\wp({x_k-x_b}, {y_k-y_b}))},
	\end{align} 
	where $\wp(x,y)=\arctan\frac{x}{y}$, $\Xi_k=f_k\cos\frac{\phi_k}{2}$, $\Xi_{k^{\prime}}=f_{k^{\prime}}\cos\frac{\phi_{k^{\prime}}}{2}$, and $d_{b,k}=\sqrt{(x_k-x_b)^2+(y_k-y_b)^2}$ is the distance between the TR and the $k$-th RE, with $(x_k,y_k)$ and $(x_b,y_b)$ representing the positions of the $k$-th RE and the TR, respectively.
\end{Proposition}
\begin{IEEEproof}
Please see Appendix B.	
\end{IEEEproof}	

From Proposition \ref{pro3}, it is observed that both DoD $\theta$ and distance $d_{0,k}$ from the target to each RE are obtained by utilizing the available knowledge, and thus the concern mentioned in Remark \ref{remark3.1} is addressed in this section. 

%
Therefore, the position of the target estimated by the $k$-th RE is given as 
\begin{align}
	\hat x_0 &=x_k+{d}_{0,k}\cos(\phi_{k}),\\
	\hat y_0 &=y_k+{d}_{0,k}\sin(\phi_k).
\end{align}
%

 
\section{Numerical and Simulation Results}\label{sec_6}
This section provides numerical and simulation results to validate the performance of the proposed algorithm for the considered multi-static ISAC system. 
 


 \subsection{Setup}\label{sec-6_1}
To promptly evaluate the performance of our proposed scheme, the number of REs in the consider the multi-static ISAC scenario is set as $K=10$, and larger number of REs $K\gg 10$ can also be operated in the same way.
 The TR position is set as $p_b = (0,0)$. The initial target position is set as $p_0=(20,40)$. The REs randomly deployed near the TR with a radius of $1\sim 100$m.
The maximum transmit power is $P_T = 30$dBm, 
 and the power of CSCG noise is set as $\sigma^2=\sigma_c^2=\sigma_z^2=-60 \text {dBm}
$. The Rician factor is set as $\alpha_k=0.5, \forall k\in\mathcal{K}$. The reflection factor is set as $0.6$. The balance factor is set as $\rho=0.5$. The numbers of the transmit antennas and receive antennas are tentatively set as $N_t = 10$ and $N_r = 2$.
The distance between any two adjacent antennas is set as $d_b=d_k=\frac{\lambda}{2}, \forall k\in\mathcal{K}$. The bandwidth of the considered system is set as $100$MHz. The time duration of each sample is set as $\Delta t = 0.5\times 10^{-8}$s. The number of samples is set as $M=1024$. 
 The pass loss exponent is set as $\epsilon=2.7$. Here, the transmit pulse function are set as $g_1(t)=\sqrt{{2}}\cos(\frac{\pi t}{2\Delta t})$ and $g_2(t)=\sqrt{\frac{\pi}{A}}\text{sinc}(\frac{t}{\Delta t})$, with $A = \int_{0}^{\pi}\frac{\sin^2(t)}{t^2}dt$.

      \begin{figure}[t]
	\setlength{\abovecaptionskip}{0.5cm}
	\centering
	\includegraphics[width=2.8in]{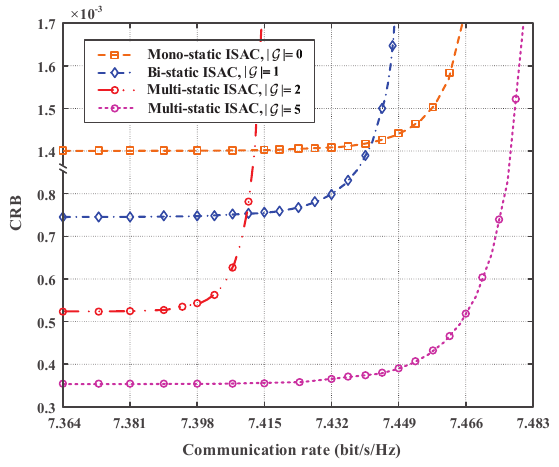}
	\caption{Trade-off between CRB and communication rate  under the mono-/bi-/multi-static ISAC scenarios.}
	\label{simulation-1}
\end{figure}
 \subsection{Performance Evaluation} 
 To study the performance trade-off between the localization and communications for our proposed scheme, the achievable region of CRB and communication rate under the mono-/bi-/multi-static scenarios are compared in Fig. \ref{simulation-1}. Here, we choose $g_1(t)=\sqrt{{2}}\cos(\frac{\pi t}{2\Delta t})$ as the transmit pulse function. 
 The mono-static ISAC scenario is set that the TR receives 
the echo with ideal SI cancellation, and the $K$ REs only receive communication signal, i.e., $|\mathcal{G}|=0$. 
 The bi-static ISAC scenario is set by choosing one RE for localization, i.e., $|\mathcal{G}|=1$. We tentatively ignore the cooperation cost, and vary the numbers of cooperative REs in our proposed multi-static ISAC scenario by setting $|\mathcal{G}|=2$ and $|\mathcal{G}|=5$, respectively. 
 From Fig. \ref{simulation-1}, it is observed that the proposed multi-static ISAC consistently outperforms both the conventional mono-static and bi-static ISAC schemes. This is due to the fact that the more ISAC RE selected, the lower the value of CRB, as shown in (\ref{equ:crb_u})-(\ref{equ:crb_u2}). Moreover, it is also revealed that the CRB under the mono-static, bi-static, and multi-static ISAC scenarios with $|\mathcal{G}|=2$ and $|\mathcal{G}|=5$ are about $1.4\times 10^{-3}$, $0.74\times 10^{-3}$, $0.52\times 10^{-3}$, and $0.36\times 10^{-3}$, respectively,  where the corresponding communication rates are $7.465$bit/s/Hz, $7.448$bit/s/Hz, $7.414$bit/s/Hz, and $7.482$bit/s/Hz, respectively.

\begin{figure}[t]
	\setlength{\abovecaptionskip}{0.5cm}
	\centering
	\includegraphics[width=2.8in]{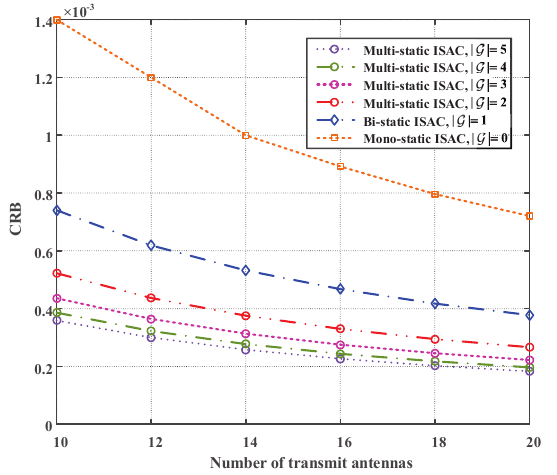}
	\caption{Effects of the number of transmit antennas on CRB for the mono-/bi-/multi-static ISAC scenarios.}
	\label{antennas1}
\end{figure}
\begin{figure}[t]
	\setlength{\abovecaptionskip}{0.5cm}
	\centering
	\includegraphics[width=2.8in]{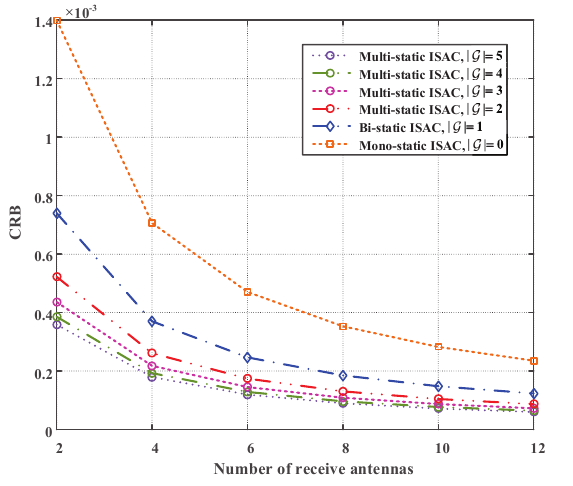}
	\caption{Effects of the number of receive antennas on CRB for the mono-/bi-/multi-static ISAC scenarios.}
	\label{antennas2}
\end{figure}

Then, the CRB of mono-/bi-/multi-static ISAC scenarios under different number of transmit antennas is plotted in Fig. \ref{antennas1}. We set the number of receive antennas at each RE as $N_r=2$, and vary the number of transmit antennas at the TR from $N_t=10$ to $N_t=20$. The transmit pulse function is set as  $g_1(t)=\sqrt{{2}}\cos(\frac{\pi t}{2\Delta t})$. It is observed that the CRB of mono-/bi-/multi-static ISAC scenarios sharply degrade with the increasing of the number of transmit antennas, e.g., the CRB under mono-static, bi-static, and multi-static ISAC scenarios with $|\mathcal{G}|=2$ decrease from $1.4\times 10^{-3}$ to $0.72\times 10^{-3}$, $0.74\times 10^{-3}$ to $0.37\times 10^{-3}$, and $0.52\times 10^{-3}$ to $0.26\times 10^{-3}$, respectively.
%
%
This is due to the fact that more transmit antennas induce more power gain. Moreover, we also vary the number of cooperative REs from $|\mathcal{G}|=2$ to $|\mathcal{G}|=5$, and it is also reveals the CRB decreasing with the number of cooperative REs increasing.

\begin{figure}[t]
	\setlength{\abovecaptionskip}{0.5cm}
	\centering
	\includegraphics[width=2.9in]{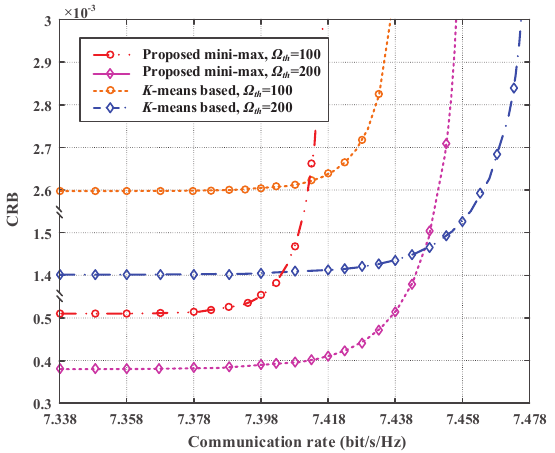}
	\caption{Trade-off between CRB and communication rate under different RE selection schemes.}
	\label{trade-off-selections}
\end{figure} 

In order to further investigate the effects of the number of receive antennas on the system performance, we plot the CRB curve with respect to different numbers of receive antennas under mono-/bi-/multi-static ISAC scenarios in Fig. \ref{antennas2}. Similarly, we set the number of transmit antennas at each RE as $N_t=10$, and vary the number of receive antennas at each RE from $N_r=2$ to $N_r=12$. 
It is revealed that the number of receive antennas has a more significant influence on the CRB compared to the number of transmit antennas, e.g., the CRB under mono-static, bi-static, and multi-static with $|\mathcal{G}|=2$ ISAC scenarios decrease from $1.4\times 10^{-3}$ to $0.24\times 10^{-3}$, $0.74\times 10^{-3}$ to $0.12\times 10^{-3}$, and $0.52\times 10^{-3}$ to $0.087\times 10^{-3}$, respectively, as shown in Fig. \ref{antennas1}. This disparity is due to the utilization of spatial diversity at the receive antennas to mitigate channel fading and noise.
Similarly, it is also shown that the CRB also decreases with the number of cooperative REs increasing.

%
%
%

Next, the trade-off between CRB and communication rate under different RE selection schemes are plotted in Fig. \ref{trade-off-selections}. Here, we set the maximum cooperation cost as $\Omega_{th} = 100$ and $\Omega_{th} = 200$, respectively. It is observed that 
 the CRB of both the proposed minimax linkage and $K$-means based methods decreases with the increasing of maximum cooperation cost, due to large number of cooperative REs. And, the CRB of the proposed minimax linkage-based RE selection outperforms that of the $K$-means based method under the same cooperation cost. 
  This is due to the fact that the conventional $K$-means based method only considers the distances among the cooperative REs, while our proposed minimax linkage-based one has incorporated the distances from the target to each RE into the objective function. Fig. \ref{trade-off-selections} also reveals that the $K$-means method achieves higher communication rate compared with the minimax linkage-based one, which is due to the fact that more power is allocated for communications in $K$-means based RE selection scheme, making poor localization performance.
  
  \begin{figure}[t]
	\setlength{\abovecaptionskip}{0.5cm}
	\centering
	\includegraphics[width=2.8in]{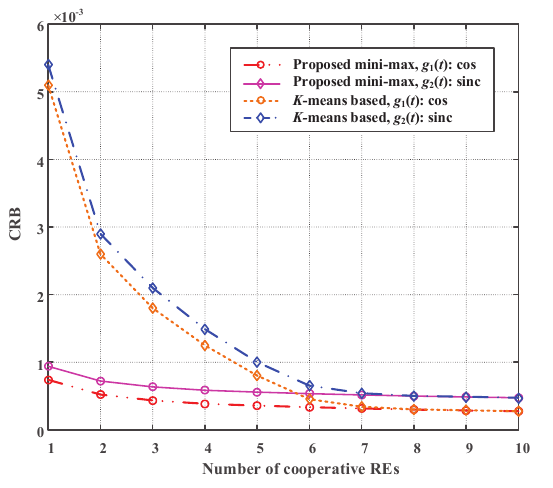}
	\caption{Effects of different numbers of cooperative REs on CRB under different transmit pulse functions and RE selection schemes.}
	\label{cooperative_cost}
\end{figure}

 Moreover, the effects of different numbers cooperative REs on CRB under different transmit pulse functions and RE selection schemes are investigated in Fig. \ref{cooperative_cost}. Here, we ignore the cooperation cost, and vary the number of cooperation REs from $1$ to $10$. It is showed that the proposed minimax linkage method achieves lower CRB than the $K$-means based method when the number of cooperation REs is small.
 However, when the number of cooperation REs approaches to $K$, both the minimax linkage and $K$-means based methods have the same localization performance. This is due to the fact that our proposed multi-static ISAC achieves the best localization performance when all the REs are cooperatively for localization, without considering the cooperation cost. We also set the transmit pulse functions as cosine and sinc in Fig. \ref{cooperative_cost}, respectively. It is revealed that the cosine pulse function always outperforms the sinc one under both minimax linkage and $K$-means based methods.

     \begin{figure}[t]
	\setlength{\abovecaptionskip}{0.5cm}
	\centering
	\includegraphics[width=2.8in]{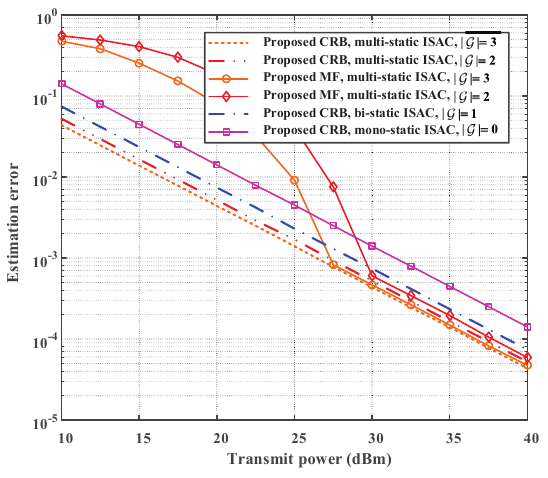}
	\caption{Comparison of proposed CRB and MF under different  power budgets.}
	\label{simulation-2}
\end{figure}

%

  Finally, the performance of MF and CRB for the joint estimations of delay and Doppler frequency shift is explicitly compared in Fig. \ref{simulation-2}, with the increasing of power budget. Here, 
  the values of $f_k$ and $\tau_k, k\in\mathcal{K},$ are obtained by (\ref{equ:f_k3}) and (\ref{equ:tau_k}) under the positions of TR, target, and REs in section \ref{sec-6_1}. Then,  the values of $\hat f_k$ and $\hat \tau_k, k\in\mathcal{K},$ are computed by (\ref{estimations}) via exhaustive grid search around the zero delay and the frequency of transmit signal \cite{1703855}. As such, the estimation error of our proposed MF in the multi-static ISAC system is computed by $\sum_{k\in\mathcal{G}}\text {Tr}({\bf M}_k^{\text{mse}})$, with ${\bf M}_k^{\text{mse}}=\mathbb{E}_{f_k,\tau_k}\Big((\hat{\boldsymbol\varrho}_k-{\boldsymbol\varrho}_k)(\hat{\boldsymbol\varrho}_k-{\boldsymbol\varrho}_k)^T\Big)$, and $2\leq|\mathcal{G}| \leq K$. Here, the number of cooperative REs are set as $2$ and $3$ in our considered multi-static ISAC system, and then the estimation error of MF and CRB of our proposed multi-static ISAC are compared in Fig. \ref{simulation-2}. It is showed that the estimation error of MF scheme is higher than the CRB in our proposed multi-static ISAC systems, and their performance gap decreases and approaches at high transmit power values. 
  Furthermore, it is also reveals that the proposed multi-static ISAC always outperforms the conventional mono-static and bi-static one under different  power budgets.

\section{Conclusions}
This paper considered a multi-static ISAC system with one TR, one target, and multiple REs. To improve the accuracy of target localization, multiple REs are expected to cooperate together for localization. The joint estimations of transmission delay and Doppler shift for localization were analyzed, whose corresponding CRB was also derived in a closed form.
Then, we formulated a CRB minimization problem for the considered multi-static ISAC system by designing the RE selection and transmit beamforming. A minimax linkage-based RE selection method and a successive convex approximation algorithm were proposed to solve this problem. We also proposed a practical method, which utilized the transmission delay and Doppler shift information of multiple REs, to realize the target localization from the electrodynamic perspective.
Finally, simulation results verified our analysis and revealed significant performance improvement of our proposed multi-static ISAC scheme over the conventional mono-static one with ideal SI cancellation when the number of cooperative REs is large.

\section*{Appendix A}\label{app-B}
\section*{Proof of Proposition \ref{pro2}}

Here, the first-order partial derivatives $\frac{\partial\ln f({\bf y}_k|{{\boldsymbol\varrho}_k})}{\partial{\tilde\tau_k}}$ and $\frac{\partial\ln f({\bf y}_k|{{\boldsymbol\varrho}_k})}{\partial{\tilde f_k}}$ are respectively computed as 
 \begin{align}
 	\label{equ:first-order1}&\frac{\partial\ln f({\bf y}_k|{{\boldsymbol\varrho}_k})}{\partial{\tilde\tau_k}} = -\frac{\partial({\bf y}_k-\boldsymbol\mu_k({\boldsymbol\varrho}_k))^H{\bf C}_k^{-1}({\bf y}_k-\boldsymbol\mu_k({\boldsymbol\varrho}_k))}{\partial{\tilde\tau_k}}\nonumber\\
 	&=\frac{\partial\boldsymbol\mu_k({\boldsymbol\varrho}_k)^H}{\partial{\tilde\tau_k}}{\bf C}_k^{-1}({\bf y}_k-\boldsymbol\mu_k({\boldsymbol\varrho}_k))\nonumber\\
 	&~~~~+({\bf y}_k-\boldsymbol\mu_k({\boldsymbol\varrho}_k))^H{\bf C}_k^{-1}\frac{\partial\boldsymbol\mu_k({\boldsymbol\varrho}_k)}{\partial{\tilde\tau_k}}, \\
 	\label{equ:first-order2}&\frac{\partial\ln f({\bf y}_k|{{\boldsymbol\varrho}_k})}{\partial{\tilde f_k}} =-\frac{\partial({\bf y}_k-\boldsymbol\mu_k({\boldsymbol\varrho}_k))^H{\bf C}_k^{-1}({\bf y}_k-\boldsymbol\mu_k({\boldsymbol\varrho}_k))}{\partial{\tilde f_k}}\nonumber\\
 	&=\frac{\partial\boldsymbol\mu_k({\boldsymbol\varrho}_k)^H}{\partial{\tilde f_k}}{\bf C}_k^{-1}({\bf y}_k-\boldsymbol\mu_k({\boldsymbol\varrho}_k))\nonumber\\
 	&~~~~+({\bf y}_k-\boldsymbol\mu_k({\boldsymbol\varrho}_k))^H{\bf C}_k^{-1}\frac{\partial\boldsymbol\mu_k({\boldsymbol\varrho}_k)}{\partial{\tilde f_k}}.
 \end{align}
Then, the second-order partial derivative of $\frac{\partial\ln f({\bf y}_k|{{\boldsymbol\varrho}_k})}{\partial{\tilde\tau_k}}$ with respect to $\tilde \tau_k$ is derived as 
\begin{align}
	&\frac{\partial^2\ln f({\bf y}_k|{{\boldsymbol\varrho}_k})}{\partial{\tilde\tau_k}^2}
	\label{equ:second2}=\frac{\partial}{\partial{\tilde\tau_k}}\left[\frac{\partial\boldsymbol\mu_k({\boldsymbol\varrho}_k)^H}{\partial{\tilde\tau_k}}{\bf C}_k^{-1}({\bf y}_k-\boldsymbol\mu_k({\boldsymbol\varrho}_k))\right]\nonumber\\
	&~~~~+\frac{\partial}{\partial{\tilde\tau_k}}\left[({\bf y}_k-\boldsymbol\mu_k({\boldsymbol\varrho}_k))^H{\bf C}_k^{-1}\frac{\partial\boldsymbol\mu_k({\boldsymbol\varrho}_k)}{\partial{\tilde\tau_k}}\right]\\
	\label{equ:second3}&=\frac{\partial}{\partial{\tilde\tau_k}}\left[\frac{\partial\boldsymbol\mu_k({\boldsymbol\varrho}_k)^H}{\partial{\tilde\tau_k}}\right]{\bf C}_k^{-1}({\bf y}_k-\boldsymbol\mu_k({\boldsymbol\varrho}_k))\nonumber\\
	&~~~~-\frac{\partial\boldsymbol\mu_k({\boldsymbol\varrho}_k)^H}{\partial{\tilde\tau_k}}{\bf C}_k^{-1}\frac{\partial\boldsymbol\mu_k({\boldsymbol\varrho}_k)}{\partial{\tilde\tau_k}}\nonumber\\
	&~~~~-\frac{\partial\boldsymbol\mu_k({\boldsymbol\varrho}_k)^H}{\partial{\tilde\tau_k}}{\bf C}_k^{-1}\frac{\partial\boldsymbol\mu_k({\boldsymbol\varrho}_k)}{\partial{\tilde\tau_k}}\nonumber\\
	&~~~~+({\bf y}_k-\boldsymbol\mu_k({\boldsymbol\varrho}_k))^H{\bf C}_k^{-1}\frac{\partial}{\partial{\tilde\tau_k}}\left[\frac{\partial\boldsymbol\mu_k({\boldsymbol\varrho}_k)}{\partial{\tilde\tau_k}}\right],
	\end{align}
where (\ref{equ:second2}) is obtained by the second-order partial derivative of (\ref{equ:first-order1}) with respect to $\tilde \tau_k$; 
(\ref{equ:second3}) is obtained by rearranging terms in  (\ref{equ:second2}). Then, the component ${\bf D}_{\tilde\tau_k,\tilde\tau_k}$ is derived as 
\begin{align}\label{equ:second4}
	&{\bf D}_{\tilde\tau_k,\tilde\tau_k}=-\mathbb{E}\left[\frac{\partial^2\ln f({\bf y}_k|{{\boldsymbol\varrho}_k})}{\partial{\tilde\tau_k}^2}\right]\\
	\label{equ:second5}&=-\mathbb{E}\left\{
	\begin{array}{l}
\frac{\partial}{\partial{\tilde\tau_k}}\left[\frac{\partial\boldsymbol\mu_k({\boldsymbol\varrho}_k)^H}{\partial{\tilde\tau_k}}\right]{\bf C}_k^{-1}({\bf y}_k-\boldsymbol\mu_k({\boldsymbol\varrho}_k))\\-\frac{\partial\boldsymbol\mu_k({\boldsymbol\varrho}_k)^H}{\partial{\tilde\tau_k}}{\bf C}_k^{-1}\frac{\partial\boldsymbol\mu_k({\boldsymbol\varrho}_k)}{\partial{\tilde\tau_k}}\\
-\frac{\partial\boldsymbol\mu_k({\boldsymbol\varrho}_k)^H}{\partial{\tilde\tau_k}}{\bf C}_k^{-1}\frac{\partial\boldsymbol\mu_k({\boldsymbol\varrho}_k)}{\partial{\tilde\tau_k}}\\
+({\bf y}_k-\boldsymbol\mu_k({\boldsymbol\varrho}_k))^H{\bf C}_k^{-1}\frac{\partial}{\partial{\tilde\tau_k}}\left[\frac{\partial\boldsymbol\mu_k({\boldsymbol\varrho}_k)}{\partial{\tilde\tau_k}}\right]\end{array}\right\}\\
	\label{equ:second6}&=2\mathbb{E}\left[\frac{\partial\boldsymbol\mu_k({\boldsymbol\varrho}_k)^H}{\partial{\tilde\tau_k}}{\bf C}_k^{-1}\frac{\partial\boldsymbol\mu_k({\boldsymbol\varrho}_k)}{\partial{\tilde\tau_k}}\right]\\
\label{equ:second7}&=\frac{2\eta_k}{G_k}\mathbb{E}\left\{
	\begin{array}{l}
\sum_{m\in\mathcal{M}}\frac{\partial{\bf x}^H[m-\tilde\tau_k]}{\partial \tilde\tau_k}\\
{\bf H}_{b,0,k}^H
{\bf H}_{b,0,k}
\frac{\partial {\bf x}[m-\tilde\tau_k]}{\partial \tilde\tau_k}
\\
\end{array}\right\}\\
\label{equ:second8}&=\frac{2\eta_k}{G_k}\mathbb{E}\left[
\begin{array}{l}
\textrm {vec}^H({\bf A}_k\textrm {vec}({\bf A}_k)
\end{array}\right]\\
\label{equ:second9}&=\frac{2\eta_k}{G_k}\mathbb{E}\left[
\begin{array}{l}
\textrm{Tr}\left(({\bf A}_k)^H{\bf A}_k\right)
\end{array}\right]\\
\label{equ:second10}&=\frac{2\eta_k}{G_k}\mathbb{E}\left\{
\begin{array}{l}
\textrm{Tr}\left[
\begin{array}{l}
{\bf H}_{b,0,k}\sum_{m\in\mathcal{M}}
\frac{\partial {\bf x}[m-\tilde\tau_k]}{\partial \tilde\tau_k}\\
\frac{\partial {\bf x}^H[m-\tilde\tau_k]}{\partial \tilde\tau_k}
{\bf H}_{b,0,k}^H
\end{array}\right]\\
\end{array}\right\}\\
\label{equ:second11}&=\frac{4BF_g\eta_k}{G_k}\mathbb{E}\left\{
\begin{array}{l}
	\textrm{Tr}\left[
	\begin{array}{l}
	{\bf H}_{b,0,k}^{H}{\bf H}_{b,0,k}\\
	\sum_{k\in\{0\}\cup\mathcal{K}}\sum_{m\in\mathcal{M}}\\
	{\bf W}_k{\bf s}_k[m]{\bf s}_k^H[m]{\bf W}_k^H
	\end{array}\right] \\
\end{array}\right\}\\
\label{equ:second12}&=\frac{4BMF_g\eta_k}{\sigma_c^2+\sigma_z^2}\left\{
\begin{array}{l}
	\textrm{Tr}\left(
		\begin{array}{l}{\bf H}_{b,0,k}^{H}{\bf H}_{b,0,k}\\
		\sum_{k\in\{0\}\cup\mathcal{K}}{\bf W}_k{\bf W}_k^H
		\end{array}\right)\\
	\end{array}
\right\},
\end{align}
where (\ref{equ:second4}) is obtained by the definition of ${\bf D}_{\tilde\tau_k,\tilde\tau_k}$ given in (\ref{equ:J_2}); (\ref{equ:second5}) is obtained by substituting (\ref{equ:second3}) into (\ref{equ:second4}); (\ref{equ:second6}) is obtained due to ${\bf y}_k-\boldsymbol\mu_k({\boldsymbol\varrho}_k)$ being a CSCG noise with zero mean, as discussed in (\ref{equ: sensing receive}); (\ref{equ:second7}) is obtained by substituting the definition of $\boldsymbol\mu_k({\boldsymbol\varrho}_k)$ and (\ref{angle1})-(\ref{angle2}) into (\ref{equ:second6}), and defining $G_k=(1+\alpha_k)(\sigma_c^2+\sigma_z^2)$; 
(\ref{equ:second8}) is obtained by setting ${\bf  A}_k={\bf H}_{b,0,k}\left[\frac{\partial{\bf x}[1-\tilde\tau_k]}{\partial\tilde\tau_k},\cdots,\frac{\partial{\bf x}[M-\tilde\tau_k]}{\partial\tilde\tau_k}\right]$; (\ref{equ:second9}) holds due to the fact of $\textrm {vec}^H({\bf X})\textrm{vec}({\bf X})=\textrm{Tr}({\bf X}^H{\bf X})$; (\ref{equ:second10}) is directly obtained by substituting the definitions of ${\bf A}_k$ into (\ref{equ:second9}); (\ref{equ:second11}) holds due to the fact that 
\begin{align}\label{equ:dotx}
	&\sum_{m\in\mathcal{M}}\frac{\partial {\bf x}[m-\tilde\tau_k]}{\partial \tilde\tau_k}\frac{\partial {\bf x}^H[m-\tilde\tau_k]}{\partial \tilde\tau_k}\nonumber\\
	&=\frac{1}{\Delta t}\int_{0}^{\Delta T}\dot{\bf x}(t)\dot{\bf x}^H(t)dt\\
	\label{equ:dotx1}&=\frac{1}{\Delta t}\int_{0}^{\Delta T}\begin{array}{l}
	\sum_{k\in\{0\}\cup\mathcal{K}}\sum_{m\in\mathcal{M}}
	{\bf W}_k{\bf s}_k[m]\\
	\dot g(t-(m-1)\Delta t)
	\dot g^*(t-(m-1\Delta t)\\
	{\bf s}_k^H[m]{\bf W}_k^Hdt\end{array}\\
	\label{equ:dotx2}&=\frac{1}{\Delta t}\begin{array}{l}	
\sum_{k\in\{0\}\cup\mathcal{K}}\sum_{m\in\mathcal{M}}{\bf W}_k{\bf s}_k[m]{\bf s}_k^H[m]{\bf W}_k^H\\
\int_{0}^{\Delta T}\dot g(t-(m-1)\Delta t)\dot g^*(t-(m-1\Delta t)dt \end{array}\\
	\label{equ:dotx3}&= \frac{1}{\Delta t}\begin{array}{l}		
\sum_{k\in\{0\}\cup\mathcal{K}}\sum_{m\in\mathcal{M}}{\bf W}_k{\bf s}_k[m]{\bf s}_k^H[m]{\bf W}_k^H\\
\int_{(m-1)\Delta t}^{m\Delta t}|\dot g(t-(m-1)\Delta t)|^2dt \end{array}\\
	\label{equ:dotx4}&={2BF_g}\sum_{k\in\{0\}\cup\mathcal{K}}\sum_{m\in\mathcal{M}}{\bf W}_k{\bf s}_k[m]{\bf s}_k^H[m]{\bf W}_k^H,
%
\end{align}
with $t\in[0,\Delta T], t=m\Delta t$, $\Delta t$ being small enough to approximate the sum by integration, and $\dot{\bf x}(t)=\frac{\partial{\bf x}(t)}{\partial t}$;
(\ref{equ:dotx1}) holding due to the fact that ${\bf x}(t)=\sum_{k\in\{0\}\cup\mathcal{K}}{\bf W}_k\sum_{m\in\mathcal{M}}{\bf s}_{k}[m]g(t-(m-1)\Delta t)$, with ${\bf s}_k(t)=\sum_{m\in\mathcal{M}}{\bf s}_{k}[m]g(t-(m-1)\Delta t)$, $\dot g(t)=\frac{\partial g(t)}{\partial t}$, and $g(t)$ being the transmit pulse function satisfying $\frac{1}{\Delta t}{\int_0^{\Delta t}|{g}(t)|^2dt}=1$; (\ref{equ:dotx2}) being obtained by rearranging the terms in (\ref{equ:dotx1}); (\ref{equ:dotx3}) being directly obtained by rearranging terms in (\ref{equ:dotx2}); (\ref{equ:dotx4}) holding by defining $F_g=\int_{(m-1)\Delta t}^{m\Delta t}|\dot g(t-(m-1)\Delta t)|^2dt=\int_{0}^{\Delta t}|\dot g(t)|^2dt$, and $\Delta t=\frac{1}{2B}$, with $F_g=\frac{\int_0^{\Delta t}|\dot{g}(t)|^2dt}{\int_0^{\Delta t}|{g}(t)|^2dt}$ being the mean square bandwidth of signal ${g}(t)$, and $B$ being the bandwidth of transmit signal. Then, (\ref{equ:second12}) is obtained by substituting ${\bf H}_{b,0,k}^{\textrm{NLOS}}$ into (\ref{equ:second11}), $G_k=\sigma_c^2+\sigma_z^2$, and ${\bf s}_k\sim\mathcal{CN}(0, {\bf I}_L)$.
From (\ref{equ:second4})-(\ref{equ:dotx4}), the proof of (\ref{equ:D_tt}) is completed.

On the other hand, the second-order partial derivative of $\frac{\partial\ln f({\bf y}_k|{{\boldsymbol\varrho}_k})}{\partial{\tilde f_k}}$ with respect to $\tilde \tau_k$ is derived as 
\begin{align}
	&\frac{\partial^2\ln f({\bf y}_k|{{\boldsymbol\varrho}_k})}{\partial{\tilde \tau_k}{\partial \tilde f_k}}
	\label{tf:second1}=\frac{\partial}{\partial{\tilde\tau_k}}\left[\frac{\partial\boldsymbol\mu_k({\boldsymbol\varrho}_k)^H}{\partial{\tilde f_k}}\right]{\bf C}_k^{-1}({\bf y}_k-\boldsymbol\mu_k({\boldsymbol\varrho}_k))\nonumber\\
	&~~~~-\frac{\partial\boldsymbol\mu_k({\boldsymbol\varrho}_k)^H}{\partial{\tilde f_k}}{\bf C}_k^{-1}\frac{\partial\boldsymbol\mu_k({\boldsymbol\varrho}_k)}{\partial{\tilde\tau_k}}-\frac{\partial\boldsymbol\mu_k({\boldsymbol\varrho}_k)^H}{\partial{\tilde\tau_k}}{\bf C}_k^{-1}\frac{\partial\boldsymbol\mu_k({\boldsymbol\varrho}_k)}{\partial{\tilde f_k}}\nonumber\\
	&~~~~+({\bf y}_k-\boldsymbol\mu_k({\boldsymbol\varrho}_k))^H{\bf C}_k^{-1}\frac{\partial}{\partial{\tilde\tau_k}}\left[\frac{\partial\boldsymbol\mu_k({\boldsymbol\varrho}_k)}{\partial{\tilde f_k}}\right],
\end{align}
where (\ref{tf:second1}) is obtained in a similar way to 
(\ref{equ:second3}). Then, the component ${\bf D}_{\tilde\tau_k,\tilde f_k}$ in matrix ${\bf J}(\tilde{\boldsymbol\varrho}_k)$ is derived as 
\begin{align}
	&{\bf D}_{\tilde \tau_k, \tilde f_k}
	\label{tf:second7}=2\mathbb{E}\left\{\mathrm{Re}\left[\frac{\partial\boldsymbol\mu_k({\boldsymbol\varrho}_k)^H}{\partial{\tilde \tau_k}}{\bf C}_k^{-1}\frac{\partial\boldsymbol\mu_k({\boldsymbol\varrho}_k)}{\partial{\tilde f_k}}\right]\right\}\\
\label{tf:second10}&=\frac{4\pi\eta_k}{G_k}\mathbb{E}\left\{\mathrm{Re}\left\{
	\begin{array}{l}
\textrm{Tr}\left[
\begin{array}{l}
{\bf H}_{b,0,k}^H{\bf H}_{b,0,k}\\
\sum_{m\in\mathcal{M}}m{{\bf x}[m-\tilde\tau_k]}\\\frac{\partial {\bf x}^H[m-\tilde\tau_k]}{\partial \tilde\tau_k}
\end{array}\right]
\end{array}\right\}\right\}\\
\label{tf:second99}&=\frac{16\pi B^2MF_{t\dot g^*}\eta_k}{(1+\alpha_k)(\sigma_c^2+\sigma_z^2)}\left\{
\begin{array}{l}
\textrm{Tr}
\left(
\begin{array}{l}
{\bf H}_{b,0,k}^H{\bf H}_{b,0,k}\\
\sum_{k\in\{0\}\cup\mathcal{K}}{\bf W}_k{\bf W}_k^H
\end{array}\right)
\end{array}
\right\},
\end{align}
where 
(\ref{tf:second7}) and (\ref{tf:second10}) are obtained in a similar way to (\ref{equ:second6}) and (\ref{equ:second11}), respectively;
(\ref{tf:second99}) holds due to
$G_k=(1+\alpha_k)(\sigma_c^2+\sigma_z^2)$ and the fact that
\begin{align}\label{equ:zero_proof}
	&\sum_{m\in\mathcal{M}}m{{\bf x}[m-\tilde\tau_k]}\frac{\partial {\bf x}^H[m-\tilde\tau_k]}{\partial \tilde\tau_k}\nonumber\\
	&=\frac{1}{\Delta t^2}\int_{0}^{\Delta T}t{\bf x}(t)\dot{\bf x}^H(t)dt\\
	\label{equ:zero_proof1}&=\frac{1}{\Delta t^2}\int_{0}^{\Delta T}
	\begin{array}{l}
	t\sum_{k\in\{0\}\cup\mathcal{K}}\sum_{m\in\mathcal{M}}{\bf W}_k{\bf s}_{k}[m]
	\\g(t-(m-1)\Delta t)\dot g^*(t-(m-1)\Delta t)\\
	{\bf s}_{k}^H[m]{\bf W}_k^Hdt
	\end{array}\\
	\label{equ:zero_proof2}&=\frac{1}{\Delta t^2}\begin{array}{l}	
\sum_{k\in\{0\}\cup\mathcal{K}}\sum_{m\in\mathcal{M}}{\bf W}_k{\bf s}_{k}[m]{\bf s}_{k}^H[m]{\bf W}_k^H\\
\int_{0}^{\Delta t}tg(t)\dot g^*(t)dt
\end{array}\\
	\label{equ:zero_proof3}&=4B^2F_{t\dot g^*}\sum_{k\in\{0\}\cup\mathcal{K}}{\bf W}_k{\bf s}_{k}[m]{\bf s}_{k}^H[m]{\bf W}_k^H, 
\end{align}
with (\ref{equ:zero_proof})-(\ref{equ:zero_proof2}) being obtained in a similar to (\ref{equ:dotx})-(\ref{equ:dotx3}), and (\ref{equ:zero_proof3}) being obtained by defining $F_{t\dot g^*}={\mathrm{Re}}\left\{\int_{0}^{\Delta t}tg(t)\dot g^*(t)dt\right\}$ \cite{923295} and ${\Delta t}=\frac{1}{2B}$. From (\ref{tf:second7})-(\ref{equ:zero_proof3}), the proof of (\ref{equ:D_tf}) is completed.
%
%

Similarly, the other two components ${\bf D}_{\tilde f_k,\tilde f_k}$ and  ${\bf D}_{\tilde \tau_k,\tilde f_k}$ in matrix ${\bf J}(\tilde{\boldsymbol\varrho}_k)$ can be proved in the same way as ${\bf D}_{\tilde \tau_k,\tilde \tau_k}$ in (\ref{equ:second4})-(\ref{equ:second12}) and ${\bf D}_{\tilde f_k,\tilde \tau_k}$ in (\ref{tf:second7})-(\ref{tf:second99}), respectively. 

Based on the above analysis, the proof of Proposition \ref{pro2} has been completed.

\section*{Appendix B}\label{app-C}
\section*{Proof of Proposition \ref{pro3}}
Based on the definition of Doppler frequency shift $f_k$ in (\ref{equ:f_k1}), the ratio of $f_k$ and $f_{k^{\prime}}, \forall k,k^{\prime}\in\mathcal{K}\backslash 0$ and $k\neq k^{\prime}$, is derived by \cite{9731802}
\begin{align}\label{equ:ratio}
	\frac{f_k}{f_{k^{\prime}}}=\frac{\cos\left(\frac{\theta+\phi_k}{2}\right)\cos\left(\frac{\theta-\phi_k}{2}\right)}{\cos\left(\frac{\theta+\phi_{k^{\prime}}}{2}\right)\cos\left(\frac{\theta-\phi_{k^{\prime}}}{2}\right)}.
\end{align}
Then, by setting $\Delta\phi=\phi_{k}-\phi_{k^{\prime}}$, (\ref{equ:ratio}) is rewritten as 
\begin{align}\label{equ:ratio2}
\frac{f_k\cos\left(\frac{\theta-\phi_{k^{\prime}}}{2}\right)}{f_{k^{\prime}}\cos\left(\frac{\theta-\phi_k}{2}\right)}=\frac{\cos\left(\frac{\theta+\phi_k}{2}\right)}{\cos\left(\frac{\theta+\phi_{k}-\Delta\phi}{2}\right)}.
\end{align}
Also, we further set $\Xi_k=f_k\cos\left(\frac{\theta-\phi_{k^{\prime}}}{2}\right), \Xi_{k^{\prime}}=f_{k^{\prime}}\cos\left(\frac{\theta-\phi_k}{2}\right)$, and $\varpi=\frac{\theta+\phi_k}{2}$, (\ref{equ:ratio2}) is equivalently transformed into
\begin{align}\label{equ:ratio3}
\frac{\Xi_k}{\Xi_{k^{\prime}}}=\frac{\cos\varpi}{\cos\left(\varpi-\frac{\Delta\phi}{2}\right)}.
\end{align}
According to the well known rule $\cos(A-B)=\cos A \cos B + \sin A \sin B$, (\ref{equ:ratio3}) is further equivalent to 
\begin{align}\label{equ:ratio4}
	\Xi_k\left(\cos\varpi\cos\frac{\Delta\phi}{2}+\sin\varpi\sin\frac{\Delta\phi}{2}\right)=\Xi_{k^{\prime}}\cos\varpi.
\end{align}
Then, dividing (\ref{equ:ratio4}) by $\cos \varpi$ and rearranging these terms, it is obtained that
\begin{align}\label{equ:ratio5}
\tan\varpi=\frac{\Xi_{k^{\prime}}-\Xi_{k}\cos\frac{\Delta\phi}{2}}{\Xi_{k}\sin\frac{\Delta\phi}{2}}.
\end{align}
From (\ref{equ:ratio5}), $\varpi$ is calculated as
\begin{align}
\varpi=\wp\left(\Xi_{k^{\prime}}-\Xi_{k}\cos\frac{\Delta\phi}{2}, \Xi_{k}\sin\frac{\Delta\phi}{2}\right),
\end{align}
where $\wp(x,y)=\arctan\frac{x}{y}$ is the four-quadrant inverse tangent function. 
Since $\varpi=\frac{\theta+\phi_k}{2}$ has been defined in (\ref{equ:ratio2}), the closed-form expression of DoA $\theta$ is calculated as
	\begin{align}\label{equ:theta}
		\theta =2\wp\left(
		\begin{array}{l}
		\Xi_{k^{\prime}}-\Xi_{k}\cos\left(\frac{\phi_{k}-\phi_{k^{\prime}}}{2}\right),\\
		\Xi_k\sin\left(\frac{\phi_{k}-\phi_{k^{\prime}}}{2}\right)
		\end{array}\right)-\phi_k.
	\end{align} 

Based on the well-known Law of Cosine in the triangle formed by the target, the TR, and the $k$-th RE, there must be
 \begin{align}\label{2d0k}
 d_{0,k}^2&=d_{b,0}^2+d_{b,k}^2-2d_{b,0}d_{b,k}\cos(\theta-\vartheta_k)\\
 \label{3d0k}&=(c\tau_k-d_{0,k})^2+d_{b,k}^2\nonumber\\
&~~~~-2(c\tau_k-d_{0,k})d_{b,k}\cos(\theta-\vartheta_k)\\
 \label{4d0k}&=c^2\tau_k^2+d_{0,k}^2-2c\tau_k d_{0,k}+d_{b,k}^2\nonumber\\
 &~~~~+2d_{0,k}d_{b,k}\cos(\theta-\vartheta_k)\nonumber\\
 &~~~~-2c\tau_k d_{b,k}\cos(\theta-\vartheta_k),
 \end{align}
 where $\vartheta_k=\wp({x_k-x_b},{y_k-y_b})$, and $d_{b,k}= \sqrt{(x_k-x_b)^2+(y_k-y_b)^2}$ is the distance between the TR and the $k$-th RE, with $(x_k,y_k)$ and $(x_b,y_b)$ being locations of the $k$-th RE and TR, respectively; (\ref{3d0k}) is obtained by substituting (\ref{equ:tau_k}) into (\ref{2d0k}); (\ref{4d0k}) is obtained by rearranging the terms in (\ref{3d0k}). Then, we have
 \begin{align}\label{1d0k}
 &2c\tau_k d_{0,k}-2d_{0,k}d_{b,k}\cos(\theta-\alpha_k)\nonumber\\
 &=c^2\tau_k^2+d_{b,k}^2-2c\tau_k d_{b,k}\cos(\theta-\vartheta_k).
 \end{align}
From (\ref{1d0k}), the distance between the target and the $k$-th RE is calculated as
\begin{align}\label{equ:d0k11}
d_{0,k}&=\frac{c^2\tau_k^2+d_{b,k}^2-2c\tau_k d_{b,k}\cos(\theta-\vartheta_k)}{2c\tau_k-2d_{b,k}\cos(\theta-\vartheta_k)}\\
\label{equ:d0k12}&=\frac{c^2\tau_k^2-2c\tau_k d_{b,k}\cos(\theta-\wp({x_k-x_b}, {y_k-y_b}))}{2c\tau_k-2d_{b,k}\cos(\theta-\wp({x_k-x_b}, {y_k-y_b}))},
\end{align}
where (\ref{equ:d0k11}) is directly from (\ref{1d0k}), and (\ref{equ:d0k12}) is obtained by substituting the definitions of $\alpha_k$ and $d_{b,k}$ into (\ref{equ:d0k11}).

From (\ref{equ:ratio})-(\ref{1d0k}), the proof of Proposition \ref{pro3} has been completed.
\bibliographystyle{IEEEtran}
\bibliography{ref}
\end{document}